\documentclass[pra,aps,floatfix,10pt,twocolumn,nofootinbib]{revtex4-1}

\usepackage[T1]{fontenc}
\usepackage{lmodern}
\usepackage[utf8]{inputenc}
\usepackage{microtype}
\usepackage{array}
\usepackage{graphicx}
\usepackage{dcolumn}
\usepackage{bm}
\usepackage{amsmath}
\usepackage{amsfonts}
\usepackage{amssymb}
\usepackage{amstext}   
\usepackage{amsthm}
\usepackage{array}
\usepackage[english]{babel}
\usepackage{makecell}
\usepackage{dsfont}
\usepackage{nameref}
\usepackage{color}
\usepackage{float}
\usepackage[outdir=./]{epstopdf}
\usepackage{nameref}
\usepackage{subfloat}
\usepackage[colorlinks = true, linkcolor=blue, urlcolor=blue, citecolor = blue]{hyperref}
\usepackage[figure]{hypcap}
\usepackage{multirow}
\usepackage{makecell}
\usepackage{xcolor}
\usepackage{mathtools}
\usepackage{mathdots}
\usepackage[notransparent]{svg}
\usepackage{todonotes} 
\newcolumntype{C}{>{$}c<{$}}
\AtBeginDocument{
\heavyrulewidth=.08em
\lightrulewidth=.05em
\cmidrulewidth=.03em
\belowrulesep=.65ex
\belowbottomsep=0pt
\aboverulesep=.4ex
\abovetopsep=0pt
\cmidrulesep=\doublerulesep
\cmidrulekern=.5em
\defaultaddspace=.5em
\tabcolsep=7pt
}
\usepackage{booktabs}

\definecolor{emerald}{rgb}{0.07, 0.53, 0.03}

\newcommand{\minskew}[1]{\mathcal{E}^{(#1)}_{\mathbf{1}}}
\newcommand{\fullmat}[1]{\mathbf{#1}}
\usepackage{amsbsy}
\newcommand{\bs}[1]{\boldsymbol{#1}}
\newcommand{\compactmat}[1]{\bs{\mathcal{#1}}}

\newcommand{\pauliX}{X}
\newcommand{\pauliY}{Y}
\newcommand{\pauliZ}{Z}

\newtheorem{theorem}{Theorem}

\begin{document}

\title{Free-Fermion Subsystem Codes}
\author{Adrian Chapman}
\email{adrian.chapman@materials.ox.ac.uk}
\affiliation{Department of Materials, University of Oxford, Parks Road, Oxford OX1 3PH, United Kingdom} 
\author{Steven T. Flammia}
\affiliation{AWS Center for Quantum Computing, Pasadena, CA 91125, USA}
\affiliation{IQIM, California Institute of Technology, Pasadena, CA 91125, USA}
\author{Alicia J.  Koll\'ar}
\affiliation{Department of Physics and Joint Quantum Institute, University of Maryland/NIST, College Park, MD 20742, USA}

\preprint{APS/123-QED}

\date{January 18, 2022}

\begin{abstract}
We consider quantum error-correcting subsystem codes whose gauge generators realize a translation-invariant, free-fermion-solvable spin model.
In this setting, errors are suppressed by a Hamiltonian whose terms are the gauge generators of the code and whose exact spectrum and eigenstates can be found via a generalized Jordan-Wigner transformation. 
Such solutions are characterized by the \emph{frustration graph} of the Hamiltonian: the graph whose vertices are Hamiltonian terms, which are neighboring if the terms anticommute.
We provide methods for embedding a given frustration graph in the anticommutation relations of a spin model and present the first known example of an exactly solvable spin model with a two-dimensional free-fermion description and exact topological qubits. 
This model can be viewed as a free-fermionized version of the two-dimensional Bacon-Shor code.
Using graph-theoretic tools to study the unit cell, we give an efficient algorithm for deciding if a given translation-invariant spin model is solvable, and explicitly construct the solution. 
Further, we examine the energetics of these exactly solvable models from the graph-theoretic perspective and show that the relevant gaps of the spin model correspond to known graph-theoretic quantities: the \emph{skew energy} and the \emph{median eigenvalue} of an oriented graph. 
Finally, we numerically search for models which have large spectral gaps above the ground state spin configuration and thus exhibit particularly robust thermal suppression of errors. 
These results suggest that optimal models will have low dimensionality and odd coordination numbers, and that the primary limit to energetic error suppression is the skew energy difference between different symmetry sectors rather than single-particle excitations of the free fermions.
\end{abstract}
\maketitle

\section{Introduction}

The rich physical behavior displayed by quantum systems makes them promising as candidate resources for quantum computational tasks.
Unfortunately, the complexity of simulating these systems makes it difficult to identify the ideal materials that can feasibly be  realized experimentally.
Exactly solvable methods provide one potential route to circumventing the difficulty imposed by these competing demands.
A particularly elegant class of systems are those which are exactly solvable via a mapping to free fermions~\cite{jordan1928uber, kitaev2006anyons}. 
Owing to this exact solution method, one can efficiently describe the eigenstates and energies of these systems classically, and additionally gain an intuitive picture for the correlations they can exhibit.

Free-fermion dynamics have a rich connection to classical and quantum complexity theory~\cite{terhal2002classical, knill2001fermionic, valiant2002quantum, bravyi2002fermionic, bravyi2006universal, jozsa2008matchgates, melo2013power, brod2014computational}, as well as to quantum error correction~\cite{kitaev2001unpaired, bravyi2010majorana, hastings2017small,  vijay2017quantum, viyuela2019scalable}. 
The latter goal is hindered in-part due to the difficulty of generating interesting and useful examples of error correcting codes that naturally relate to free-fermion systems.
Recently however, there has been some progress in systematically \emph{recognizing} free-fermion-solvable spin models~\cite{chapman2020characterization, ogura2020geometric}.

In this paper, we leverage the graph-theoretic tools of Ref.~\cite{chapman2020characterization} to solve the ``inverse'' problem of embedding a free-fermion system into a spin model in such a way as to generate useful subsystem codes.
This is closely related to the problem of finding good fermion-to-qubit mappings~\cite{verstraete2005mapping, setia2018superfast, seeley2012bravyikitaev, bravyi2017tapering, steudtner2018fermion, jiang2019optimal, derby2021compact, chiew2021optimal,  jiang2019majorana}, where one seeks to find a qubit model whose dynamics --- possibly over a restricted subspace --- are equivalent to a desired interacting fermion model.
Here, we are treating the effective fermionic degrees of freedom as the gauge qubits of a subsystem code, rather than encoding our logical quantum information.
In that sense, the models which we consider are perhaps more akin to the 1-d Kitaev wire~\cite{kitaev2001unpaired}. 
This and related models~\cite{bravyi2010majorana} are desirable for protection against symmetry-respecting noise.

Our formalism is additionally inspired by a recent characterization given by Haah~\cite{haah2013commuting} for compactly describing translation-invariant error-correcting stabilizer codes using Laurent polynomials, which we adapt to the setting of graphs for our purposes.

We present the first known example of a topological subsystem code that can be mapped to a two-dimensional free-fermion model and which contains exact logical qubits.
These logical degrees of freedom are topologically protected and have no analog in the effective free-fermion model. 
In contrast, prior comparable models lacked at least one of these ingredients, and either had non-exact string operators~\cite{kitaev2006anyons}, or they were a union of 1D (or even 0D) free-fermion models~\cite{Yu2008Topological, bravyi2012subsystem}, or they were not free-fermion solvable~\cite{Bacon2006, Yu2008Topological, Bombin2010Topological, bravyi2010majorana, suchara2011constructions}.
This model is explicitly constructed so as to be free-fermion solvable. 
For the case of translation-invariant spin models which are obtained by other means, we present an efficient recognition algorithm for detecting whether or not these models are free-fermion solvable. If the model is exactly solvable, our algorithm automatically constructs the corresponding root graph occupied by the constituent free-fermions.

In addition to this algorithm, we extend previous binary-valued linear algebraic descriptions of translation-invariant spin models and error-correcting codes~\cite{haah2013commuting} to include a description for all relevant quantities: the frustration graph of the spin model, the root graph on which the free-fermion solution lives, the spin model itself, and calculation of the free-fermion energies.

We examine the energetics of possible free-fermion solutions, and show that the relevant spin-model gaps which control energetic suppression of errors correspond to graph-theoretic quantities, namely the recently introduced skew energy~\cite{adiga2010skew} and the median eigenvalues of orientations of the root graph. 
Finally, we carry out numerical studies of these gaps for a large family of one and two-dimensional test lattices and identify design heuristics for finding candidate models with large gaps based on both dimensionality and coordination number. 
The results of this numerical search indicate that the primary energetic bottleneck to intrinsic error suppression in the known cases is the skew energy gap between orientations of the root graph, rather than the median eigenvalue which corresponds to the single-particle gap of a free-fermion model at half filling.

This paper is organized as follows. 
After reviewing the background of our formalism in Sec.~\ref{sec:background}, our main results are given from Sec.~\ref{sec:examples} onward.
In Sec.~\ref{subsec:honeycombandfiducial}, we develop tools for realizing a given frustration graph, and we discuss two methods in particular: the honeycomb bosonization and the fiducial bosonization, which are used to construct two examples of exactly solvable spin models with exact logical qubits. The first example, in Sec.~\ref{subsec:doublyfreefermion}, is the first known example of a free-fermion model with a two-dimensional frustration graph exhibiting exact logical qubits. 
The second example, in Sec.~\ref{subsec:trianglemodel}, illustrates a potential pathological case in which a free-fermion-solvable model and exact logical qubits coexist but are unrelated.

In Sec.~\ref{sec:lgrecognize}, we give an algorithm which produces the compact Laurent-polynomial description of a root graph given the corresponding description of its translation-invariant line graph. 
In Sec.~\ref{sec:laurentspinmodels}, we present a generalized binary-valued matrix encoding which extends the compact Laurent-polynomial formalism to include the original spin model in addition to the root and frustration graphs.
In Sec.~\ref{sec:skewEconnections}, we show the connection between the ground-state energies of free-fermion models and the graph-theoretic notions of the \emph{skew energy} and median eigenvalue of a graph.
Having identified these connections we examine their implications for finding free-fermion models with exact logical qubits and large gaps, which can intrinsically suppress errors.
In Sec.~\ref{sec:numerics}, we numerically calculate the relevant energy gaps in a series of example free-fermion solvable spin models, and demonstrate and identify empirical heuristics which describe the trade-off between properties of these graphs and properties of the underlying free-fermion models.

\section{Background}
\label{sec:background}
\subsection{Free-Fermion-Solvable Spin Models}
\label{sec:ffbackground}
We consider a many-body spin model defined on $n$ qubits with Hamiltonian given by
\begin{align}
    H_{s} = \sum_{\boldsymbol{j} \in E} h_{\boldsymbol{j}} \sigma^{\boldsymbol{j}} \mathrm{.}
    \label{eq:spinmodel}
\end{align}
Here, $\sigma^{\boldsymbol{j}} \in \mathcal{P}_n$ denotes an $n$-qubit Pauli operator, and the coupling coefficients $h_{\boldsymbol{j}}$ are necessarily all real. 
Paulis are labeled by bit strings $\boldsymbol{j} \in \{0, 1\}^{\times 2n}$ as $\sigma^{\boldsymbol{j}} = i^{|\boldsymbol{j_x}\cdot\boldsymbol{j_z}|} X^{\boldsymbol{j_x}}Z^{\boldsymbol{j_z}}$, where $\boldsymbol{j} = \boldsymbol{j}_x \boldsymbol{j}_z$ is the concatenation of two $n$-bit strings $\boldsymbol{j_x}$ and $\boldsymbol{j_z}$.
The sum above runs over a set $E$, which is just the set of strings $\boldsymbol{j}$ where $h_{\boldsymbol{j}}$ is nonzero. 
A useful quantity is the \emph{symplectic binary form} $\langle \cdot, \cdot\rangle$, defined as
\begin{align}\label{eqn:steveencoding}
    \langle\boldsymbol{j},\boldsymbol{k}\rangle = \boldsymbol{j_x}\cdot\boldsymbol{k_z} + \boldsymbol{j_z}\cdot\boldsymbol{k_x} \bmod 2\,.
\end{align}
The commutator of two $n$-qubit Paulis is given by
\begin{align}
    \sigma^{\boldsymbol{j}} \sigma^{\boldsymbol{k}} = (-1)^{\langle \boldsymbol{j},\boldsymbol{k}\rangle} \sigma^{\boldsymbol{k}} \sigma^{\boldsymbol{j}} \,.
\end{align}

We will focus on those spin models of the form above which can be exactly solved by a mapping to free fermions.
A free-fermion model has a Hamiltonian of the form
\begin{align}
    H_f = i \sum_{(j, k) \in \widetilde{E}} h_{jk} \gamma_j \gamma_k \equiv \frac{i}{2} \boldsymbol{\Gamma}^{\mathrm{T}} \cdot \mathbf{h} \cdot \boldsymbol{\Gamma} \mathrm{.}
\label{eq:ffdefinition}
\end{align}
The column vector $\mathbf{\Gamma} \equiv \left(\gamma_j \right)_{j \in V}$ consists of the \emph{Majorana operators}, which satisfy the canonical anticommutation relations
\begin{align}
    \{\gamma_{j}, \gamma_{k}\} \equiv \gamma_{j}\gamma_{k} + \gamma_{k}\gamma_{j} = 2\delta_{jk}I \mathrm{.}
    \label{eq:car}
\end{align}
Like Pauli operators, products of Majorana operators only commute or anticommute with each other and square to $\pm I$. 
The Majorana operators themselves are Hermitian. 
The sets $V$ and $\widetilde{E}$ above simply label the distinct Majorana operators and the nonzero coupling terms respectively, but their names foreshadow their use in the graph-theoretic formalism that we will introduce shortly. 

The matrix of coupling coefficients, $\mathbf{h} \in \mathds{R}^{|V| \times |V|}$, is called the \emph{single particle Hamiltonian}. 
Hermiticity of $H_f$ and the $\{\gamma_j\}$, together with Eq.~(\ref{eq:car}), imply that we may take $\mathbf{h}$ to be an antisymmetric matrix: $\mathbf{h}^{\mathrm{T}} = -\mathbf{h}$. 
We can thus associate $\mathbf{h}$ to a directed graph with orientation $\tau$, $R^{(\tau)} = (V, \widetilde{E}^{(\tau)})$, such that an arc $(j \rightarrow k) \in \widetilde{E}^{(\tau)}$ if $h_{jk} > 0$.
Where necessary, we will correspondingly label the relevant operators by the orientation $\tau$.

An exact solution for $H_f$ can be found by block-diagonalizing the single-particle Hamiltonian
\begin{align}
    \mathbf{h} = \mathbf{w} \cdot \bigoplus_{j = 1}^{|V|/2} \begin{pmatrix}
    0 & -\lambda_j \\
    \lambda_j & 0
    \end{pmatrix} \cdot \mathbf{w}^{\mathrm{T}}
    \label{eq:orthdiag}
\end{align}
if $|V|$ is even. 
If $|V|$ is odd, $\mathbf{h}$ has an additional zero eigenvalue.
From the diagonalizing orthogonal matrix, $\mathbf{w} \in SO(|V|)$, it is straightforward to construct a unitary $W$ which preserves the canonical commutation relations
(\ref{eq:car}) and diagonalizes $H_f$
\begin{align}
    W^{\dagger} H_{f} W = -i \sum_{j = 1}^{\lfloor |V|/2 \rfloor} \lambda_{j} \gamma_{2j - 1} \gamma_{2j} \mathrm{.}
    \label{eq:diagonalferm}
\end{align}
In this basis, the terms of the Hamiltonian commute and each square to a scalar operator. 
The Hamiltonian spectrum, $\mathcal{E}_{\boldsymbol{x}}$, is thus harmonic in the Williamson eigenvalues, $\{\lambda_j\}$, of $\mathbf{h}$
\begin{align}
    \mathcal{E}_{\boldsymbol{x}} = \sum_{j = 1}^{\lfloor |V|/2\rfloor} (-1)^{x_j} \lambda_{j} \mathrm{.}
    \label{eq:ffspectrum}
\end{align}
Here, $\boldsymbol{x} \in \{0, 1\}^{\times \lfloor |V|/2 \rfloor}$ labels an eigenstate of $H_{f}$ and represents the filling configuration. 
For this reason, the $\{\lambda_j\}$ are also referred to as the single-particle energies. 
Note that the spectrum of $H_f$ is symmetric about zero.

Revisiting the spin Hamiltonian $H_s$ in Eq.~(\ref{eq:spinmodel}), this model admits a mapping to a free-fermion Hamiltonian when its Pauli terms exhibit the same pair-wise commutation relations as the Majorana terms in a Hamiltonian $H_f$ of the form in Eq.~(\ref{eq:ffdefinition}) for some undirected graph $R$. 
Define the \emph{frustration graph}, $G(H)$, of a Hamiltonian, $H$, whose interaction terms either commute or anticommute, as the graph whose vertices correspond to terms in $H$ and for which vertices are neighboring if their corresponding terms anticommute.
Given a graph $R = (V, E)$, called the root graph, its line graph, $L(R) = (E, F)$, is the graph describing the incidence relations between the edges of $R$.
That is, $(e_1, e_2) \in F$ if edges $e_1$, $e_2 \in E$ share a common vertex in $R$.

A central result of Ref.~\cite{chapman2020characterization} is the following:

\begin{theorem}[{Ref.~\cite[Thm.~1]{chapman2020characterization}}] 
Given an $n$-qubit spin Hamiltonian of the form in Eq.~(\ref{eq:spinmodel}) with frustration graph $G(H_s)$. There exists an injective map $\varphi: E \mapsto V^{\times 2}$ effecting 
\begin{align}
\sigma^{\boldsymbol{j}} \mapsto i \gamma_{\varphi_1(\boldsymbol{j})} \gamma_{\varphi_2(\boldsymbol{j})} 
\label{eq:ffmap}
\end{align}
such that 
\begin{align}
\sigma^{\boldsymbol{j}} \sigma^{\boldsymbol{k}} = (-1)^{|\varphi(\boldsymbol{j}) \cap \varphi(\boldsymbol{k})|} \sigma^{\boldsymbol{k}} \sigma^{\boldsymbol{j}}
\end{align}
if and only if there exists a root graph $R$ such that
\begin{align}
    G(H_s) \simeq L(R),
\label{eq:ffsolution}
\end{align} 
where R is the hopping graph of the free-fermion solution.
\label{eq:thm1chapman}
\end{theorem}
The above theorem gives necessary and sufficient conditions to associate a unique pair of Majorana fermions to each Pauli term in $H_s$ such that the commutation relations are preserved. 
However, this only defines the mapping $\varphi$ up to exchanges, $\varphi_1(\boldsymbol{j}) \leftrightarrow \varphi_2(\boldsymbol{j})$, corresponding to a sign freedom on the elements of $\mathbf{h}$, or an orientation on the root graph $R$.
We must specify this choice such that products of Pauli terms are preserved by the mapping as well.
In the spin picture, products of Pauli terms in $H_s$ are enforced by constraints of the form
\begin{align}
    \prod_{\boldsymbol{j} \in S} \sigma^{\boldsymbol{j}} = i^{d(S)} \sigma^{\boldsymbol{k}(S)}
    \label{eq:stabconstraints}
\end{align}
for subsets $S \subseteq E$.
Here, $\sigma^{\boldsymbol{k}(S)}$ is a symmetry (possibly the identity) that commutes with every Pauli term of $H_s$. 
Since $\varphi$ is chosen such that commutation relations are preserved, we consider similar products among the terms of $H_f$.
These products are exactly those subsets $S$ of edges in $R$ such that either (i) every vertex $\boldsymbol{j} \in V$ is incident to even-many edges in $S$ or (ii) every vertex is incident to odd-many edges in $S$. 
Products of terms from $H_f$ in case (i) are generated by cycles in $R$ and give the identity
\begin{align}
\prod_{\boldsymbol{j} \in C} \left(i \gamma_{\varphi_1(\boldsymbol{j})} \gamma_{\varphi_2(\boldsymbol{j})}\right) = (-1)^{\tau(C)} i^{|C|} I,
\label{eq:cyclephase}
\end{align}
where the product is taken over a cycle $C$ in $R$, in cyclic order.
The phase factor is determined by $|C|$, the number of edges in $C$, and $\tau(C)$, the orientation of the cycle.
The latter quantity is the number of times we need to commute $\gamma_{\varphi_1(\boldsymbol{j})}$ and $\gamma_{\varphi_2(\boldsymbol{j})}$ on the left-hand-side such that individual Majorana operators cancel pairwise.
The aforementioned case (ii) is only possible if the number of vertices in $R$ is even. 
In this case, products of terms from $H_f$ are generated by $T$-joins of $R$ (see Ref.~\cite{chapman2020characterization}) and give the parity operator
\begin{align}
\prod_{\boldsymbol{j} \in T} \left(i \gamma_{\varphi_1(\boldsymbol{j})} \gamma_{\varphi_2(\boldsymbol{j})}\right) = \pm i^{|V|/2} \prod_{j \in V} \gamma_j \mathrm{.}
\end{align}
If $\sigma^{\boldsymbol{k}(S)}$ is not the identity for some $S$, we restrict to the eigenspace stabilized by $\pm \sigma^{\boldsymbol{k}(S)}$ and solve the model by free fermions on that eigenspace by choosing the root-graph orientation to be consistent with Eq.~(\ref{eq:stabconstraints}).
As shown in Ref.~\cite{chapman2020characterization}, this can be done for any cycle-symmetry configuration.
If the parity operator gives the identity in the spin picture (up to multiplication by cycle operators), we project onto a fixed-parity subspace of our free-fermion solution.

To summarize, we can solve a spin model if its frustration graph is a line graph. 
The solution takes each of the subspaces labeled by symmetries of the model and maps that symmetry sector to a free-fermion model.
We obtain a solution over each stabilizer subspace by choosing a suitable orientation of the root graph.
Fixing the stabilizer and choosing a particular state of the free-fermion degrees of freedom may not completely specify a state in the Hilbert space, however.
There may be Pauli operators that commute with $H_s$ but that cannot be made as products of Hamiltonian terms (and so are not captured by Eq.~(\ref{eq:stabconstraints})).
These degrees of freedom will constitute logically encoded qubits for our subsystem codes.

\subsection{Quantum subsystem codes}\label{subsec:subsystemcodes}
A quantum stabilizer code is specified by an abelian subgroup $\mathcal{S} \subseteq \mathcal{P}_n$ such that $-I \notin \mathcal{S}$, whose mutual $+1$-eigenspace constitutes the logical codespace.
Because $\mathcal{S}$ is a group, it is sufficient to specify the codespace by the mutual $+1$-eigenspace of any set of generators of $\mathcal{S}$. 
The elements of $\mathcal{S}$ thus preserve, or \emph{stabilize}, the codespace.
The Pauli subgroup of operators that commute with every element in $\mathcal{S}$ is called the \emph{centralizer} of $\mathcal{S}$, denoted $\mathcal{C}(\mathcal{S})$.
The logical Pauli group for the codespace is given by $\mathcal{C}(\mathcal{S})/ \mathcal{S}$, commuting elements to $\mathcal{S}$ that are outside of $\mathcal{S}$ itself.
Since the codespace is stabilized by elements of $\mathcal{S}$, logical operators are defined up to equivalence by stabilizers.

A subsystem code~\cite{poulin2005stabilizer, nielsen2011quantum,  bravyi2012subsystem} is defined similarly, but the requirement that the elements of $\mathcal{S}$ commute is relaxed.
Rather, a subsystem code is defined by a nonabelian gauge group $\mathcal{G} \subset \mathcal{P}_n$. 
Again, logical operators are specified as elements of the centralizer $\mathcal{C}(\mathcal{G})/\mathcal{G}$. 
There is an associated stabilizer group for the subsystem code $\mathcal{S} \equiv \mathcal{C}(\mathcal{G}) \cap \mathcal{G}$.
A subsystem code can therefore be understood as a stabilizer code where the state of some of the logical qubits, the gauge qubits, can be ignored.
The gauge group constitutes the Pauli group on the gauge qubits together with the stabilizers of this code.
Crucially, by including noncommuting operators in the gauge group, we can often generate the group using lower-weight gauge generators than any stabilizer code with the same code space.

Errors from the subsystem code logical space can be suppressed by an error suppression Hamiltonian, which is a sum over a (possibly overcomplete) set of gauge generators. 
To analyze the error correction properties of the model, it is therefore germane to consider the energetics of this Hamiltonian.
When the Hamiltonian derives from a stabilizer code, the energetics is easy to analyze. 
However, for a subsystem code the terms are generally noncommuting, so solving the model is a nontrivial task. 
This is why we restrict to code Hamiltonians that are free-fermion solvable.

\subsection{Translation-Invariant Graphs}\label{subsec:LaurentLattices}

Here we summarize an algebraic construction of Haah~\cite{haah2013commuting} for describing translation-invariant graphs. 
Consider a graph $G \equiv (V, E)$ that exhibits translation-invariance in one dimension. 
Such a graph has a \emph{block Toeplitz} adjacency matrix
\begin{align}
    \mathbf{A} = \begin{pmatrix}
    \ddots & \vdots & \vdots & \vdots & \\ \dots & \mathbf{A}_0 & \mathbf{A}_1 & \mathbf{A}_2 & \dots \\
    \dots & \mathbf{A}_1^{\mathrm{T}} & \mathbf{A}_0 & \mathbf{A}_1 & \dots \\
    \dots & \mathbf{A}_2^{\mathrm{T}} & \mathbf{A}_1^{\mathrm{T}} & \mathbf{A}_0 & \dots \\
    & \vdots & \vdots & \vdots & \ddots
    \end{pmatrix}
\end{align}
where $\mathbf{A}_k \in \mathds{F}_2^{n \times n}$ describes the adjacency relations between a unit cell of $n$ vertices and the unit cell displaced by $k$ sites.

We compress this description to a finite-dimensional adjacency matrix over a ring as follows. 
We associate translation by $k$ sites to the right with the monomial term $x^k$. 
In this way, we construct the following matrix with Laurent-polynomial-valued entries as 
\begin{align}
    \mathbf{A}[x] \equiv \sum_{k = -\infty}^{\infty} \mathbf{A}_k x^k
\end{align}
with $\mathbf{A}_{-k} \equiv \mathbf{A}_k^{\mathrm{T}}$. 
It is convenient to distinguish between translation to the left and to the right by introducing a new variable $\bar{x}$ such that $x\bar{x} = 1$. 
Then the Laurent-polynomial-valued matrix can naturally be expressed as having entries in the polynomial ring $\mathbb{F}_2[x,\bar{x}]/\langle x\bar{x} - 1\rangle$.

\begin{figure}
\centering
		\includegraphics[width=0.45\textwidth]{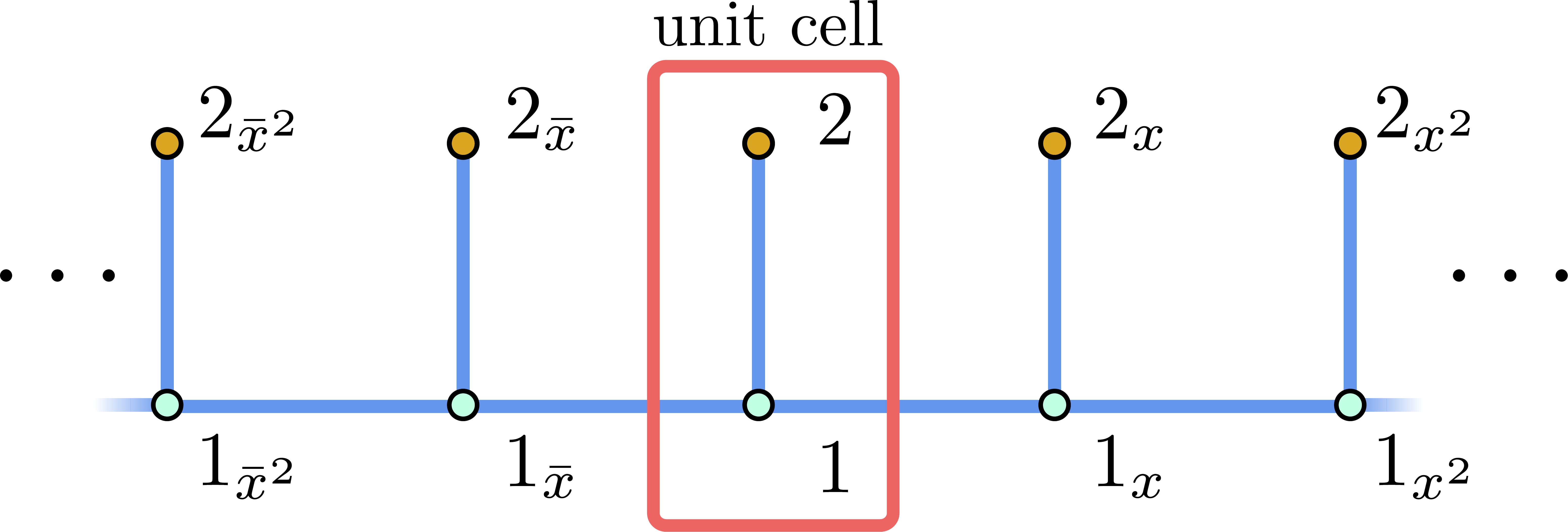}
	\caption{Translation-invariant comb graph. We denote the translation of a graphical structure by a subscript (e.g. the translation of vertex ``1'' by one unit cell in the $x$-direction by $1_x$, two unit cells by $1_{x^2}$, and so on).}
	\label{fig:combgraph}
\end{figure}

From here on, we denote polynomial-valued compact matrices with calligraphic font to distinguish them from binary-valued non-compact ones. 
For example, the $2 \times 2$ adjacency matrix
\begin{align}
    \compactmat{A} = \begin{pmatrix}
    x + \bar{x} & 1 \\
    1 & 0
    \end{pmatrix},
    \label{eq:tpathmatrix}
\end{align}
describes the infinite 1-d ``comb'' with two sites per unit cell shown in Fig.~\ref{fig:combgraph}.
The block matrices describing the connections between a unit cell and one translated by $x^\nu$ are denoted by: 
\begin{align}
    \mathbf{A}_x = \begin{pmatrix}
    1 & 0 \\
    0 & 0
    \end{pmatrix},
\end{align}
\begin{align}
    \mathbf{A}_{\bar{x}} = \begin{pmatrix}
    1 & 0 \\
    0 & 0
    \end{pmatrix},
\end{align}
and
\begin{align}
    \mathbf{A}_1 = \begin{pmatrix}
    0 & 1 \\
    1 & 0
    \end{pmatrix}.
\end{align}
We will use the same notation of subscripting by a monomial $x^{\nu}$ to indicate graph structures (e.g. vertices and cliques) translated by $\nu$ unit cells from the origin.

This can be extended to $d$ dimensions by introducing a set of variables $\boldsymbol{x} \equiv \{x_1, x_2, \dots, x_d \}$. 
Let $\mathcal{A}_{\boldsymbol{x}^{\mathbf{v}}} \in \mathds{F}_2^{n \times n}$ describe the adjacency relations between a unit cell of $n$ vertices and the unit cell displaced by a vector $\mathbf{v} \in \mathds{Z}^{\times d}$, then
\begin{align}
    \compactmat{A} &= \sum_{\mathbf{v} \in \mathds{Z}^{\times d}} \mathbf{A}_{\boldsymbol{x}^{\mathbf{v}}} \boldsymbol{x}^{\mathbf{v}}
    \label{eq:compactdef}
\end{align}
captures all graphs with translation invariance over $d$ dimensions, since any lattice can be coarse-grained to a hypercubic lattice. 
We have introduced the abridged notation
\begin{align}
\boldsymbol{x}^{\mathbf{v}} \equiv \prod_{j = 1}^d x_j^{v_j}
\end{align}

An example of this in two dimensions is the adjacency matrix for the kagome lattice, shown in the background of Fig.~\ref{fig:triangleModel}\textbf{b},
\begin{align}
    \compactmat{A}^{\mathrm{Kagome}} = \begin{pmatrix}
    0 & 1 + \bar{x} & \bar{x} + y \\
    1 + x & 0 & 1 + y \\
    x + \bar{y} & 1 + \bar{y} & 0
    \end{pmatrix}
\end{align}

\section{Examples}\label{sec:examples}
 
In the following section, we will present two motivating and illustrative examples of exactly solvable models with two-dimensional frustration graphs and logical qubits. 
Crucially, both of these examples contain both a non-commuting free-fermion-solvable Hamiltonian as well as commuting stabilizer terms; however the role that these types of terms play in the storage of quantum information is dramatically different between the two.
To the best of our knowledge, the first example, which is described in Section \ref{subsec:doublyfreefermion}, is the only known example of an exactly solvable spin model with a two-dimensional effective free-fermion solution \emph{and} a constant number of exact logical qubits with nontrivial distance. 
The second example, described in Section \ref{subsec:trianglemodel}, illustrates an undesirable trivial case in a which a free-fermion solvable model and a stabilizer code coexist, with only the latter giving rise to the logical degrees of freedom. 
This second case illustrates some of the difficulties in deciding in advance which subset of these constructions is ``useful'' in new and nontrivial ways.

\subsection{Realizing a Given Frustration Graph}\label{subsec:honeycombandfiducial}

\begin{figure*}[t!]
\centering
		\includegraphics[width=0.98\textwidth]{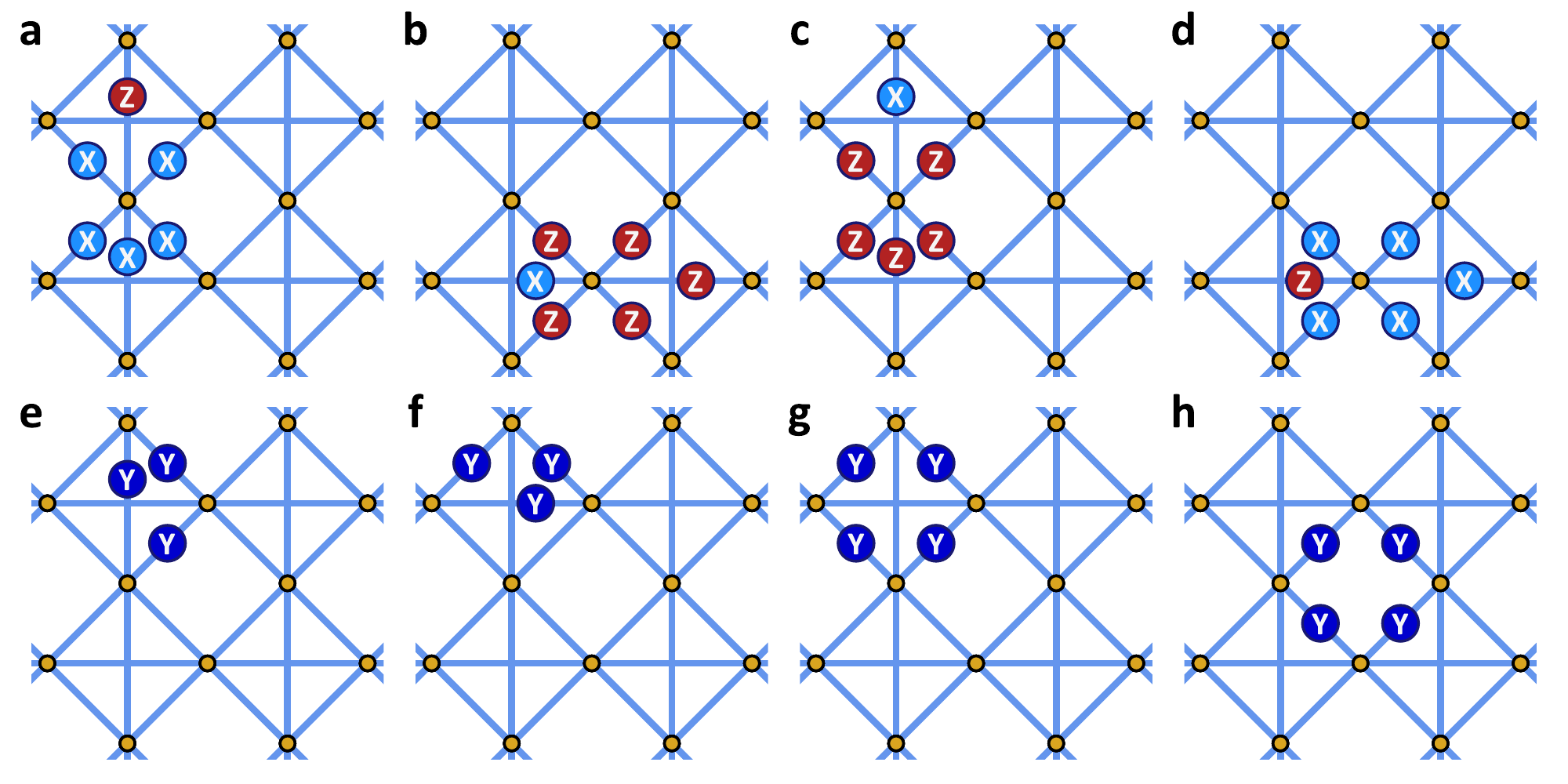}
	\caption{The terms of the checkerboard-lattice code. 
	\textbf{a} and \textbf{b} The two inequivalent Hamiltonian terms of a fiducial bosonization $H_0$ of the line graph of the square lattice. 
	The line graph is shown in light blue, with brown circles to indicate the vertices. 
	Note that there is no vertex in the middle of the square plaquettes where the horizontal and vertical edges cross. 
	The constituent single-qubit Pauli operators of each term are indicated in color-coded and labeled circles in the center of the each edge. 
	For ease of view, we indicate the identity by the absence of a label, and for the two edges that cross, we have displaced the drawn symbols to make it clear which edge they belong to. 
	\textbf{c} and \textbf{d} The two inequivalent terms of a second fiducial bosonization $H_1$ obtained by flipping all single-qubit Paulis. 
	All terms in $H_0$ commute with all terms in $H_1$.
	\textbf{e}-\textbf{h} four additional stabilizer terms per unit cell. 
	Combining both Hamiltonians and the four stabilizers gives a frustration graph which consists of two independent copies of the line graph of the square lattice and produces two exact logical qubits, shown in Fig.~\ref{fig:FreeFermionLogicals}.}
	\label{fig:doubleFreeFermion}
\end{figure*}

In order to generate the two examples in Sections \ref{subsec:doublyfreefermion} and \ref{subsec:trianglemodel} and be able to connect them to other potential models with different two-dimensional frustration graphs, we must first introduce the concept of a bosonization in order to connect a graph with potential spin models which realize it as a frustration graph. 

A \emph{bosonization} is an explicit realization of a given frustration graph by a qubit-Pauli Hamiltonian.
A full treatment of the procedure for finding and classifying all such realizations has been discussed elsewhere~\cite{tantivasadakarn2020jordanwigner, chen2018exact} and is beyond the scope of this work. 
In Section \ref{sec:laurentspinmodels}, we will discuss how to unify the description of these models with the Laurent-polynomial description of lattices in Section \ref{subsec:LaurentLattices}. 
For now, we concentrate on a separate aspect of the bosonization problem: construction of a spin model which realizes a given translation-invariant frustration graph.

The spin model which realizes a given frustration graph is by no means unique, and here we will consider only bosonizations that share the translation invariance of the frustration graph, albeit possibly with a larger unit cell. 
Even subject to this constraint, there are still many possible spin models which bosonize a given frustration graph.
In general it is desirable for the spin model to be local, or quasi-local, or consist only of very low-order terms, and solving the bosonization problem subject to such ``physicality'' and ``hardware-implementability'' constraints is beyond the scope of this work. 
Here we will consider only two extremely simple, but highly general methods, which we call the \emph{fiducial} and \emph{honeycomb} bosonizations.

\emph{\textbf{The honeycomb bosonization}} is a minor generalization of the the Kitaev honeycomb model~\cite{kitaev2006anyons} and can be applied to any frustration graph which is the line graph of a root graph with maximum degree less than or equal to three. 
In this bosonization a qubit is assigned to each \emph{vertex} of the \emph{root graph}, and each of the (half) edges emerging from the vertex is assigned to either $\pauliX$, $\pauliY$, or $\pauliZ$ of that qubit. 
An edge between vertices $\alpha$ and $\beta$ which is $\pauliX$ on the $\alpha$ side and $\pauliY$ on the $\beta$ side then becomes $\pauliX_{(\alpha)}\pauliY_{(\beta)}$. 
This bosonization consists only of low-order terms, but it can only be applied to a select set of graphs.

\emph{\textbf{The fiducial bosonization}} is more general. 
In contrast to the honeycomb bosonization, it amounts to assigning a qubit to every \emph{edge} of the \emph{frustration graph}. 
The two ends of each edge are assigned to either $\pauliX$ or $\pauliZ$, and the vertices of the frustration graph are Hamiltonian terms which are the tensor product of the different factors of $\pauliX$ and $\pauliZ$ from the corresponding ends of \emph{all} the edges incident on that vertex. 
This bosonization is fully general and can be applied to any frustration graph, including line graphs with maximum degree four or greater, but it generally results in experimentally unfriendly multiqubit Hamiltonian terms. 
Additionally, the fiducial bosonization is relatively inefficient, and in general there exists a large number of local Pauli operators that commute with a fiducial bosonization Hamiltonian.
The cycle $\pauliY$ stabilizers shown in Fig. \ref{fig:doubleFreeFermion} \textbf{e}-\textbf{h} are one example of a mutually commuting set. However, there are many others which are not supported on closed cycles, and many of these do not commute with one another.

The combination of the the honeycomb and fiducial bosonizations gives a way to produce a spin model which realizes \emph{any} frustration graph, whether it is a line graph and whether it is free-fermionizable or not, regardless of whether the frustration graph is translation invariant. 
Since these two methods guarantee that there exists at least one spin model which realizes a given frustration graph, throughout much of this paper we will examine frustration graphs without considering any specifics of the corresponding spin model.

The examples given in the next two sections are an exception. 
Here will will consider the anticommuting free-fermion Hamiltonian terms and additional stabilizers explicitly in order to show that exactly solvable spin models with a constant number of exact logical qubits and two-dimensional frustration graphs do exist. 
These example models will also illustrate some of the general phenomenology and pitfalls of these types of models, as well as these two bosonization methods in particular.

\begin{figure}[t!]
\centering
		\includegraphics[width=0.48\textwidth]{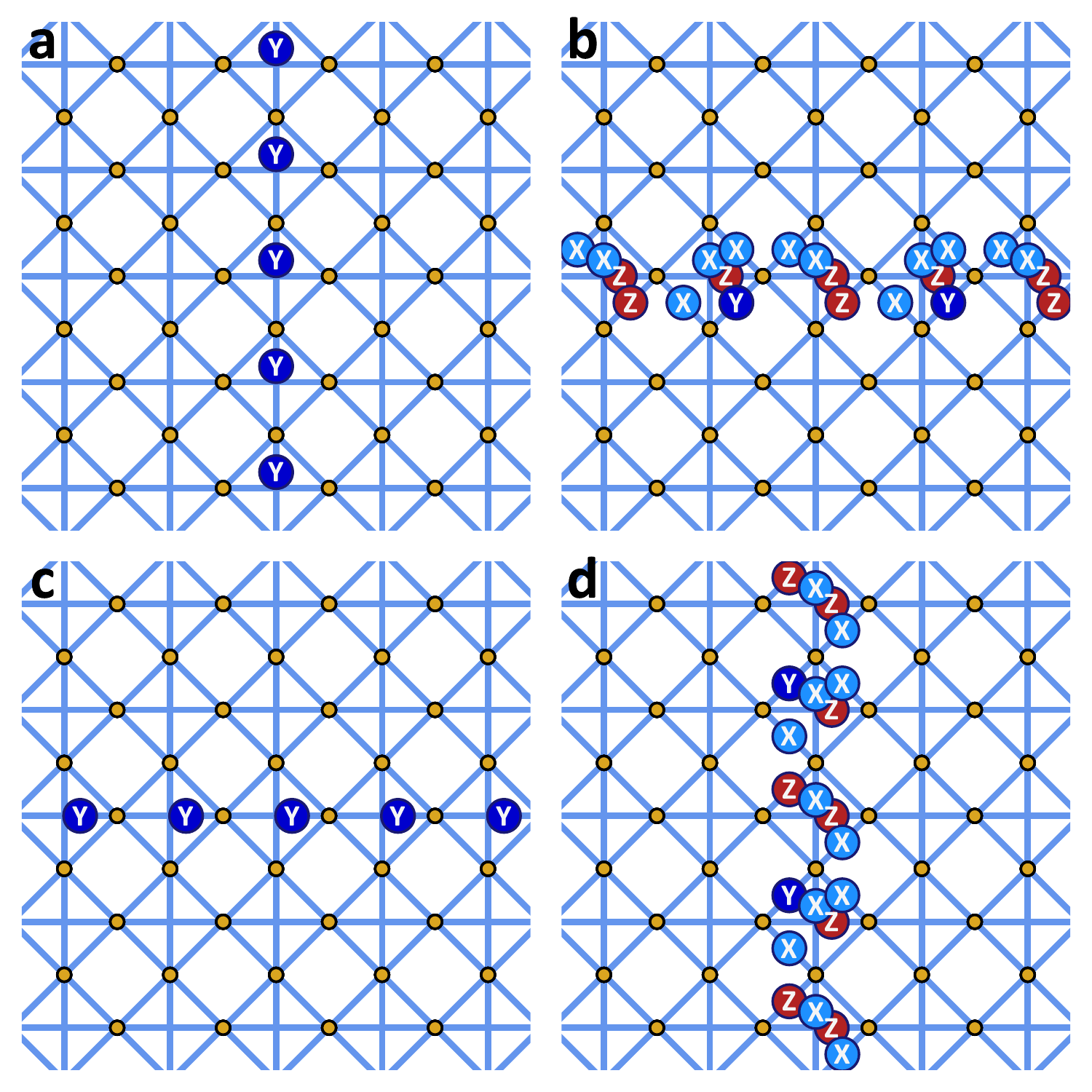}
	\caption{Logical degrees of freedom of the checkerboard-lattice code. 
	\textbf{a} and \textbf{b} String-like logical $\pauliX$ and logical $\pauliZ$ operators that form an exact logical qubit which commutes with all the Hamiltonian and stabilizer terms shown in Fig.~\ref{fig:doubleFreeFermion}. \textbf{c} and \textbf{d} A second string-like logical pair which commutes with the first. 
	On a torus with even dimensions, these are the only independent operators that commute with all the Hamiltonian and stabilizer terms. \label{fig:FreeFermionLogicals}}
\end{figure}

\subsection{The Checkerboard-Lattice Code: Intertwined Free-Fermion Models} \label{subsec:doublyfreefermion}

The first model we present is based on a fiducial bosonization of the line graph of the square lattice combined with an additional set of stabilizers. 
The line graph of the square lattice is a checkerboard lattice in which every other plaquette has diagonal (next-nearest-neighbor) edges, shown in the background of all the subfigures of Fig.~\ref{fig:doubleFreeFermion}. 
Since this is a $6$-regular graph, all fiducial bosonization terms are weight 6. 
There are two inequivalent Hamiltonian terms, one for the horizontal edges of the square lattice, and one for the vertical ones. 
The simplest possible orientation and realization of the Hamiltonian terms is shown in Fig.~\ref{fig:doubleFreeFermion}\textbf{a} and \textbf{b}. 
Note that the horizontal and vertical edges of the line graph connect equivalent sites in neighboring unit cells, so it is not possible to chose the Hamiltonian terms to consist of only one type of single-qubit Pauli operator. 
By construction, the fiducial Hamiltonian operators in Fig.~\ref{fig:doubleFreeFermion}\textbf{a} and \textbf{b} and all of their translates have a frustration graph which is the line graph of the square lattice, and the resulting Hamiltonian can be solved exactly by computing a free-fermion model on the square lattice. 

Because all sites in this line graph have even degree, reversing the choice of orientation on all edges gives rise to a second fiducial bosonization, shown in Fig.~\ref{fig:doubleFreeFermion}\textbf{c} and \textbf{d}, such that all terms the new fiducial Hamiltonian $H_1$ commute with all terms in the original Hamiltonian $H_0$. 
As a result, the total Hamiltonian to $H_t = H_0 + H_1$ has a frustration graph which consists of two disconnected copies of the line graph of the square lattice, and is still free-fermion solvable. 
In addition to the two free-fermion Hamiltonians, we will also include $4$ stabilizers per unit cell, shown in Fig.~\ref{fig:doubleFreeFermion}\textbf{e}-\textbf{h}, which consist of products of $\pauliY$ around closed cycles. 
In addition to these stabilizers and their linear combinations, there is another operator per unit cell that commutes with all of the Hamiltonian terms which is given by the product of Hamiltonian terms (either \textbf{a} and \textbf{b} or \textbf{c} and \textbf{d}, but not mixed combinations) around a plaquette of the underlying square lattice. 
The two loop operators formed this way are equivalent up to products of the $\pauliY$ stabilizers and generated by the Hamiltonian terms, so they are not independent.

Under $L \times L$ periodic boundary conditions with $L$ even, there are precisely four linearly independent operators that commute with all the stabilizers and both free-fermion Hamiltonians. 
They form two exact logical qubit pairs shown in Fig.~\ref{fig:FreeFermionLogicals}. 
One half of each logical pair is an incontractable loop of the $\pauliY$ stabilizers and consists of $\pauliY$ along the vertical or horizontal edges. 
We will refer to these operators as the $Y$ strings. 
The other halves of each logical pair, which we call the $XZ$ strings, are more complicated operators related to the free-fermion Hamiltonian terms. 
These operators, shown in Fig.~\ref{fig:FreeFermionLogicals}\textbf{b} and \textbf{d}, break the discrete translation symmetry of the underlying lattice and repeat only every two unit cells. 
The $XZ$ string which is translated by one unit cell still commutes with all of the Hamiltonian and stabilizer terms, but will anticommute with the string operator running around the torus the other way. 

The two $XZ$ strings in each direction can be considered to be a factorization of a product of Hamiltonian terms along the same line. 
For example, the product of the $XZ$ string shown in Fig.~\ref{fig:FreeFermionLogicals}\textbf{b} with its translated partner and the $Y$ string in Fig.~\ref{fig:FreeFermionLogicals}\textbf{c} is equal to the product of the Hamiltonian term in Fig.~\ref{fig:doubleFreeFermion}\textbf{d} along the same line. 
Thus, the logical degrees of freedom arise because homologically non-trivial products of Hamiltonian terms can be broken up into string operators which still commute with all Hamiltonian and stabilizer terms individually. 
Unlike, e.g.~string operators in the Kitaev honeycomb model, these fractions of incontractable products of Hamiltonian terms are linearly independent of the Hamiltonian terms. 

Another way to view the free-fermion Hamiltonian terms of this model is by analogy to the Bacon-Shor code~\cite{Shor1995,Bacon2006,NappPreskill2013}, and the one-dimensional $XY$ chain. 
The Hamiltonian terms can be thought of as the products of two terms: $\pauliX \pauliX \pauliX \pauliX$ or $\pauliZ \pauliZ \pauliZ \pauliZ$ on the edges at $45^\circ$ and $\pauliX \pauliZ$ on the vertical or horizontal edges. 
The quad $\pauliX$ or quad $\pauliZ$ operators resemble vertex stabilizers of the toric code~\cite{Kitaev:2003toric}. 
However, the alternation between $\pauliX$ and $\pauliZ$ at every other vertex of the line graph produces anticommuting terms whose frustration graph is the square lattice, the same as that of the two-dimensional Bacon-Shor code. 
Additionally, the second fiducial bosonization contains all of the vertex terms of the opposite type, so combining the two includes $\pauliX \pauliX \pauliX \pauliX$ and $\pauliZ \pauliZ \pauliZ \pauliZ$ at \emph{every} site of the line graph.

A subsystem code made from the toric-code-like operators alone would be a variation on the two-dimensional Bacon-Shor code, but since the square lattice is not a line graph, it would not be free-fermion solvable by the methods discussed in this work \footnote{While there are alternative methods available to find an exact free-fermion solution (see e.g.~\cite{elman2021free}), the Bacon-Shor code is not amenable to these, as its frustration graph contains claws and even holes.}. 
The necessary missing edges are introduced by the $\pauliX \pauliZ$ operators on the vertical and horizontal edges. 
Taken by themselves, these operators give rise to models and frustration graphs which are equivalent to disconnected copies of the one-dimensional $\pauliX \pauliY$ chain running around the torus in both directions.
The one-dimensional chains have the same frustration graph as the 1-d Kitaev wire, with the important distinctions that they are periodic. For us, the logical operators have no analog in terms of fermion operators.

Since the fiducial Hamiltonian terms are products of these two types of generators, acting on completely separate qubits, their frustration graph is the two square lattices from the vertex-like operators combined with one-dimensional chains in the vertical and horizontal direction. 
This combination produces the exactly solvable checkerboard lattice from the non-free-fermion square-lattice frustration graph of the two-dimensional Bacon-Shor code.

In sum, this model gives an example of a completely local spin model for which exact logical degrees of freedom arise not from stabilizers or a free-fermion model alone, but from a combination of the two. 
The frustration graph of the Hamiltonian is two-dimensional, albeit not connected, consisting of two copies of the line graph of the square lattice. 
In contrast, prior comparable models lacked at least one of these ingredients, and either had non-exact string operators~\cite{kitaev2006anyons}, or they were a union of 1D (or even 0D) free-fermion models~\cite{Yu2008Topological, bravyi2012subsystem}, or they were not free-fermion solvable~\cite{Bacon2006, Yu2008Topological, Bombin2010Topological, bravyi2010majorana, suchara2011constructions}

The full energy spectrum of the system can be found by analyzing the three constituent components independently: the $\pauliY$ loop stabilizers, and the two \emph{independent} free-fermion models on separate copies of the square lattice. 
Furthermore, the ground state orientation for both free-fermion halves, and therefore the model as a whole, is known from Ref.~\cite{lieb1994flux} even though it cannot be determined from recent graph-theoretic results~\cite{adiga2010skew}.
As yet, however, we know of no example of a free-fermion solvable model with exact logical degrees of freedom that does not involve a significant set of commuting stabilizers.

\subsection{Triangle Models}\label{subsec:trianglemodel}

\begin{figure*}[t!]
\centering
		\includegraphics[width=0.98\textwidth]{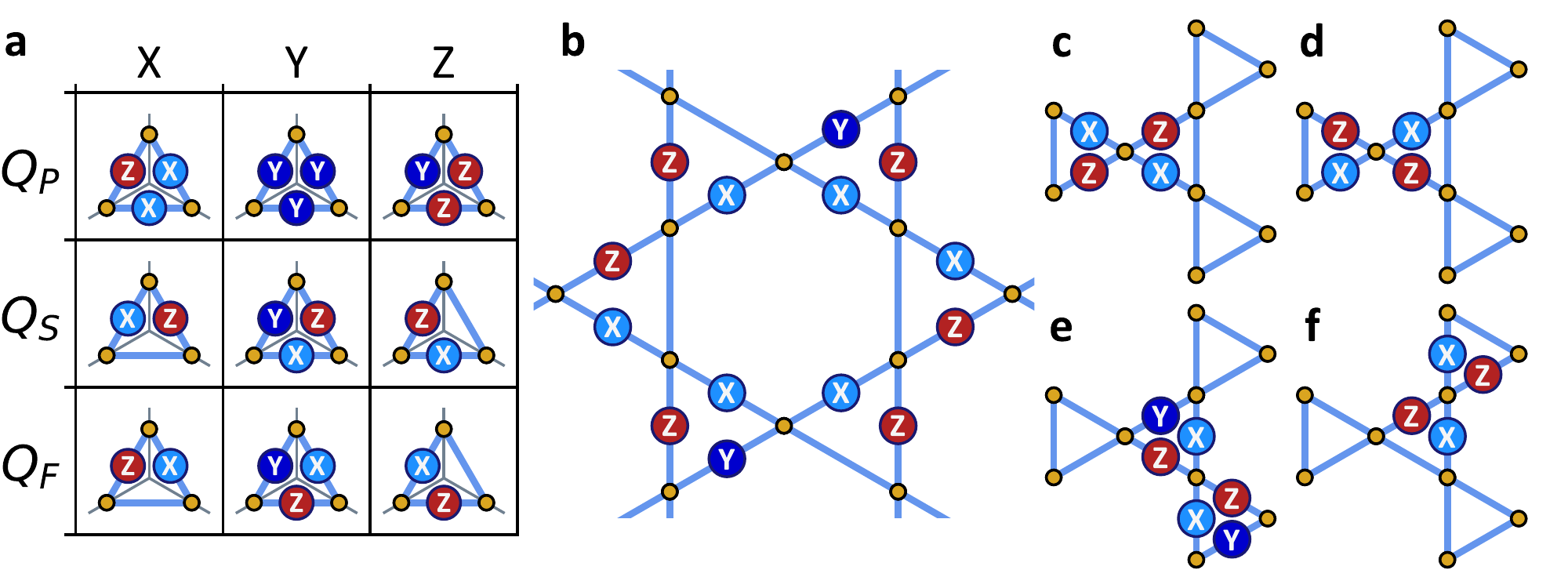}
	\caption{Building blocks of a triangle model. \textbf{a} Table showing the three effective qubits $Q_P$, $Q_S$, and $Q_F$ formed from three physical qubits on the edges of a triangular plaquette. Physical-qubit Pauli operators are indicated in color-coded and labeled circles in the center of the each edge. For ease of view, we indicate the identity by the absence of a label. \textbf{b}, \textbf{c} Stabilizers of a Wen plaquette model constructed from the effective qubit $Q_S$. The loop stabilizer is the product $\pauliX^{(S)} \pauliY^{(S)} \pauliZ^{(S)} \pauliX^{(S)} \pauliY^{(S)} \pauliZ^{(S)}$, and the bond stabilizer is $\pauliZ^{(S)}\pauliZ^{(S)}$. Note that the definition of $\pauliZ^{(S)}$ is rotated by 180 degrees on the two inequivalent plaquettes of the $ZZ$ bond. \textbf{d}-\textbf{f} $ZZ$, $YY$, and $XX$ bonds, respectively, of a Kitaev honeycomb model defined using all the translates of $Q_F$. Despite being defined on the same set of physical qubits, every term in the Kitaev honeycomb model in $Q_F$ commutes with every term of the Wen plaquette model in $Q_S$, and all of the Paulis of $Q_P$.}
	\label{fig:triangleModel}
\end{figure*}

The second example we present is a construction based on line graphs with triangular plaquettes which allows a conventional stabilizer code to be combined with a free-fermion-solvable model on the same set of physical qubits. 
Because of the heavy reliance on having triangular plaquettes, we refer to these models as triangle models. 
We will present the particular case of a Wen plaquette model combined with a Kitaev honeycomb model, but the construction is more general. 
We highlight this particular example because it shows two general properties of free-fermion-solvable spin models. 
The first notable feature is that while this model has free-fermion character and gives rise to topological logical-qubit degrees of freedom, the quantum information storage is due to commuting terms which do not enter the free-fermion model. 
Our particular construction is designed to make this separation explicit. 
However, if a free-fermion-solvable model with topological logical qubits is found through some other means, the existence of a such a separation may not be obvious by inspection, and care would need to be taken to distinguish this trivial case from a model in which the free-fermion degrees of freedom truly store quantum information. 
The second relevant aspect of this model is that it shows that for $3$-regular graphs, the fiducial bosonization is closely related to a honeycomb model constructed from a set of effective qubits. 

The fundamental building block of this example is to place qubits on the edges of the line graph $L(R)$ of a $3$-regular graph $R$ and define commuting effective qubits on the triangular plaquettes of $L(R)$. 
We will first present the general construction of these triangle qubits before moving to the explicit model under consideration.

Given a $3$-regular graph $R$, its line graph $L(R)$ consists entirely of triangular plaquettes which share at most one corner. 
For our particular example, we will let $R$ be the hexagonal honeycomb and $L(R)$ the kagome lattice, shown in Fig.~\ref{fig:triangleModel}\textbf{b}. 
We then assign three physical qubits to each plaquette, one on each edge. 
From these physical qubits, we can define three effective qubits, which we denote by $Q_P$, $Q_S$, and $Q_F$. 
The form of the individual effective Pauli operators in terms of the physical qubits is shown in Fig.~\ref{fig:triangleModel}\textbf{b}. 
There is one copy of each of these qubits for every triangular \emph{plaquette} in $L(R)$, i.e.~for every \emph{vertex} in $R$. 

Since Pauli operators of the three effective qubits are independent and commute with one another, we can view the resulting Hilbert space as being a model defined on three independent copies of $R$, and we are free to choose any desired Hamiltonian for each different effective qubit. 
For a concrete example, we use the $Q_S$ qubits to build a Wen plaquette model, the $Q_F$ qubits to build a Kitaev honeycomb model, and we stabilize the remaining $Q_P$ degrees of freedom by treating it as a paramagnet in an effective magnetic field long the $y$ direction. 

In order to make the Hamiltonian terms as simple and local as possible in terms of the physical qubits, it is necessary to select a judicious orientation of the two inequivalent triangular plaquettes in each unit cell of the kagome lattice and a matching assignment of each pair of corner-sharing triangles to $\pauliX$, $\pauliY$, or $\pauliZ$ bonds.
Since we have three independent types of qubits, we will denote the corresponding Pauli operators by a superscript of their corresponding type, e.g.~$\pauliZ^{(S)}$. The operators $\pauliX^{(S)}$ and $\pauliZ^{(S)}$ cluster around one vertex of the triangular plaquette, so it is natural to associate them to this corner and the underlying edge of $R$. 
The operator $\pauliY^{(S)}$ is then naturally associated to the remaining corner/edge, and $Q_F$ and $Q_P$ inherit the same association. 
For this specific example, we chose to orient the two inequivalent triangles such that the $\pauliZ$ operators correspond to the horizontal edges of $R$. 
The resulting $\pauliZ^{(S)}\pauliZ^{(S)}$ term of the Wen plaquette model is shown in Fig.~\ref{fig:triangleModel}\textbf{c}. 
All of the remaining terms follow from maintaining this orientation in every unit cell. 
Figure~\ref{fig:triangleModel}\textbf{d} shows the $\pauliZ^{(F)}\pauliZ^{(F)}$ term of the Kitaev honeycomb model, and Fig.~\ref{fig:triangleModel}\textbf{e} and \textbf{f} the $\pauliY^{(F)}\pauliY^{(F)}$ and $\pauliX^{(F)}\pauliX^{(F)}$ terms, respectively. 
The final two Hamiltonian/stabilizer terms required are $\pauliY^{(P)}$ on every plaquette (shown in Fig.~\ref{fig:triangleModel}\textbf{a}) and the loop operator of the Wen plaquette model $\pauliX^{(S)} \pauliY^{(S)} \pauliZ^{(S)} \pauliX^{(S)} \pauliY^{(S)} \pauliZ^{(S)}$ (shown in Fig.~\ref{fig:triangleModel}\textbf{b}).

In sum, this model consists of three independent spin models defined on the same set of physical qubits: a stabilizer code which stores quantum information, a free-fermion solvable model, and paramagnet to pin down the remaining degrees of freedom. 
Weight-one Pauli operators are smaller than the effective qubits, and will therefore anticommute with at least one Hamiltonian term from both the stabilizer-code and the free-fermion model, thereby endowing the stabilizer code with some of the energetics of the free-fermion model. 
However, any Pauli made entirely of $\pauliX^{(S)}$, $\pauliY^{(S)}$, and $\pauliZ^{(S)}$ will commute with the free-fermion Hamiltonian. 
It is thus clear that the combined model still suffers from the low-energy string errors that plague 2D stabilizer codes, and that the free-fermion model is ancillary to the quantum information storage.

As a result, we draw a general conclusion that if a Hamiltonian with topological logical qubits contains both a free-fermion-solvable model and a set of stabilizer terms that commute with each other and the free-fermion model, then the free-fermion terms need not play any role in the quantum information storage.
The triangle-model construction makes this separation very explicit, but in other models, with more complex effective qubits, the existence of independent constituent models may not be readily apparent. 
An exceedingly simple example of this arises when the operator $\pauliY^{(F)}$ is redefined to also include a factor of $\pauliY^{(P)}$. 
This local rotation mixes the paramagnet qubit $Q_P$ into the free-fermion qubit $Q_F$, but does not affect the commutation relations of \emph{any} of the Hamiltonian terms. 
It also has the effect of exactly transforming the Kitaev honeycomb model terms (shown in Fig.~\ref{fig:triangleModel}\textbf{d}-\textbf{f}) into a fiducial bosonization on the kagome lattice.

Since the triangle-model construction only relied on the existence of triangular plaquettes which share at most one corner, the following relation between the Kitaev honeycomb model and fiducial bosonizations emerges. 
Given a $3$-regular graph $R$, a fiducial bosonization of $L(R)$ is equivalent to a honeycomb bosonization of $R$, up to local rotations. 
Furthermore, the vast number of operators that commute with the fiducial bosonization can be understood as coming from the (largely) independent triangle qubits which are not involved in the honeycomb bosonization. 

For graphs of degree greater than three, $L(R)$ no longer consists of corner-sharing triangles, so the close correspondence to a honeycomb model on $R$ breaks down. 
However, the ``triangle'' construction can in fact be extended to non-regular graphs whose \emph{maximum} degrees is three. 
We will not present the construction here, because the resulting models are qualitatively the same, and handling the more general case requires addressing special cases and the possible addition of extra local stabilizers for lower-degree vertices. 
However, we have verified that starting from a graph which is $3,1$-biregular, it is possible to make a triangle model in which the stabilizer code portion is a generalized toric code with three qubits per edge instead of one.

\section{Line-Graph Recognition}

To determine if a translation-invariant spin Hamiltonian has a free-fermion solution using Theorem 1 of Ref.~\cite{chapman2020characterization}, we must be able to recognize that a given translation-invariant graph is a line graph. 
To do this, we utilize the Whitney isomorphism theorem~\cite{whitney1932congruent}, which says that if two connected graphs with more than four vertices are edge-isomorphic, then there exists exactly one vertex isomorphism that induces the edge isomorphism, with all of the small exceptions known. 
A \emph{Krausz decomposition}~\cite{krausz1943demonstration} of a line graph $L(R) \equiv (E, F)$ is a partition of its edges in $F$ into cliques (complete subgraphs) $\{K^{(1)}, K^{(2)}, \dots, K^{(|V|)}\}$ such that every vertex in $E$ belongs to at most two of the subgraphs induced by the $\{K^{(i)}\}$. 
If we allow vertices to belong to cliques containing no edges, then we can define the decomposition such that every vertex in $E$ belongs to exactly two cliques. 
Under the line graph operation $L$, each vertex $v \in V$ in the root graph $R \equiv (V, E)$ is mapped bijectively to the clique $K^{(v)}$. 
Whitney isomorphism guarantees that for large enough graphs, this decomposition is unique.

\subsection{Recognition Algorithm}\label{subsec:lgalgorithm}

\label{sec:lgrecognize}

\begin{figure}
\centering
		\includegraphics[width=0.45\textwidth]{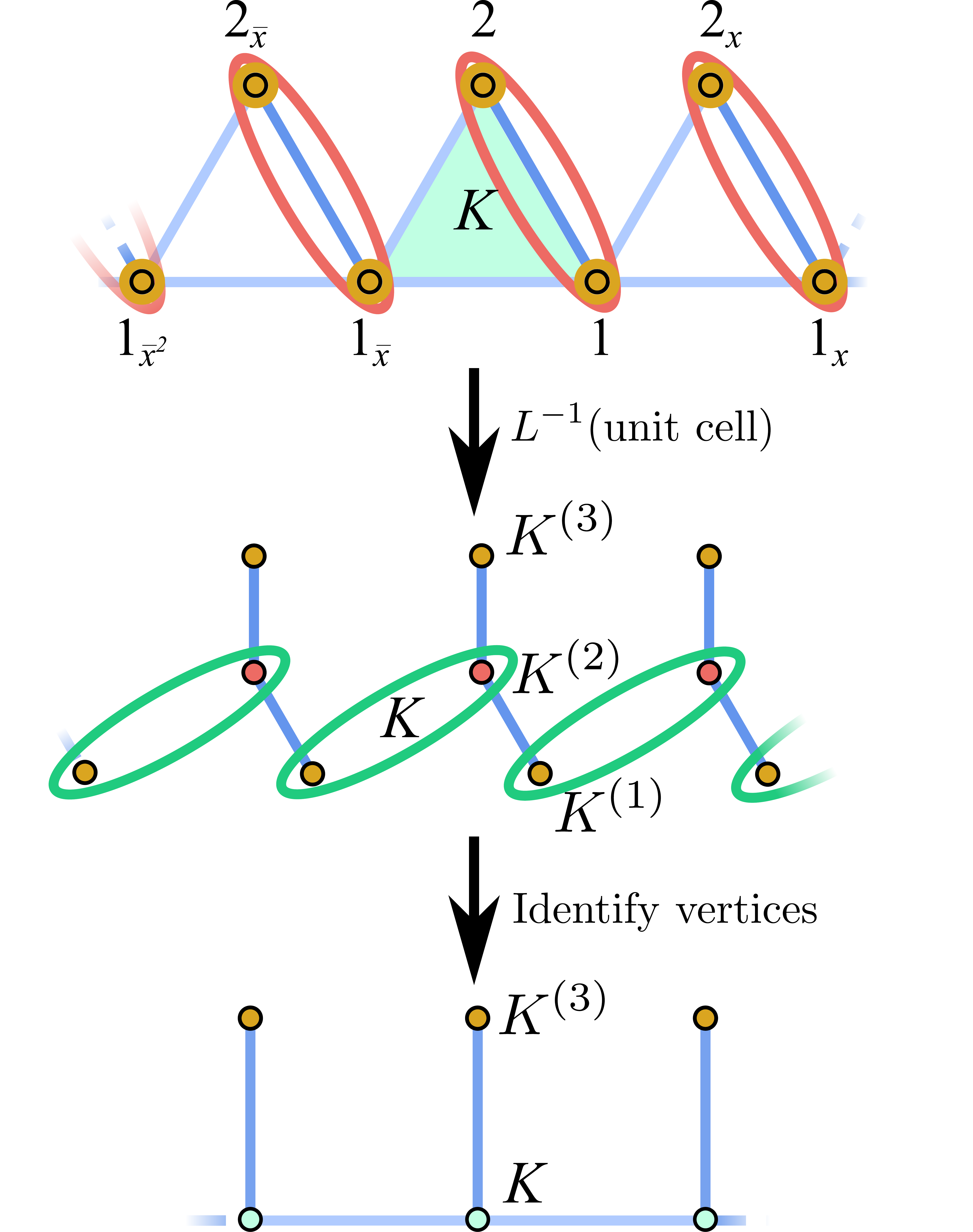}
	\caption{An illustration of our line-graph recognition algorithm for translation-invariant graphs. 
	(Top) The ``triangle path graph.'' 
	We first perform line-graph recognition on the individual unit cells, partitioning edges into cliques (circled) such that every vertex belongs to exactly two cliques. 
	(Middle) The root graphs of individual unit cells.
	We next identify vertices according to the connections between unit cells. 
	In this case, we identify $K^{(1)}$ with $K^{(2)}_{x}$ to recover the clique $K$ shown in the top figure.
	(Bottom) After identification, we obtain the correct root graph.}
	\label{fig:trianglepath}
\end{figure}

The first step is to examine the subgraph induced by the vertices in a single unit cell. 
This is performed by looking at the constant term in the polynomial expansion. 
For the 1-d example shown in Fig.~\ref{fig:trianglepath}, we have
\begin{align}
    \mathbf{A}^{\triangle-\mathrm{path}}_1 = \begin{pmatrix}
    0 & 1 \\
    1 & 0
    \end{pmatrix}
\end{align}
We next take the Krausz decomposition of this subgraph. 
If the decomposition does not exist, then the global, translation-invariant graph cannot be a line graph, and we are done. 
If the decomposition is not unique, we can coarse-grain the lattice, say along one direction, until the unit cell has more than five vertices. 
Though we should really think of the cliques in this decomposition as bins of edges, we designate each clique by the vertices it contains, as the edges in the clique are completely determined by this. 
For our 1-d example, we have 
\begin{align}
    K^{(1)} = \{1\}\mathrm{,} \mbox{\hspace{5mm}} K^{(2)} = \{1, 2\}\mathrm{,} \mbox{\hspace{5mm}} K^{(3)} = \{2\} \mathrm{,} 
\end{align}
where cliques containing one vertex are understood to contain no edges. 
We finally include the edges not in these cliques by verifying that they are consistent with the global Krausz decomposition. 
Once again, we will denote an object (vertex, edge, or clique) in a translated unit cell by its index in the ``origin'' unit cell with its translation-monomial as a subscript. 
For example, $1_x$ denotes the vertex $1$ in the unit cell translated from the origin by $x$, and $K^{(1)}_y$ denotes the clique $K^{(1)}$ in the unit cell translated from the origin by $y$.

Note that the edges between unit cells cannot induce any new cliques in the Krausz decomposition, since every vertex already belongs to two cliques (including empty ones), and there are no vertices not in any clique. 
However, it may be necessary to identify (i.e.~take the union of) cliques between unit cells. 
To do this, we check if the edges corresponding to each distinct monomial induce sets of disconnected complete bipartite graphs between cliques in their different cells, and these are the cliques we identify. 
If one clique, say $K^{(1)}$, is identified with more than one clique, say $K^{(2)}_x$ and $K^{(3)}_y$, then we must also be able to identify $K^{(2)}$ with $K^{(3)}_{\bar{x}y}$ (i.e.~the three cliques, $K^{(1)}$, $K^{(2)}$, and $K^{(3)}$ are all identified).

For our 1-d example, we have
\begin{align}
 \mathbf{A}^{\triangle-\mathrm{path}}_x = \begin{pmatrix}
    1 & 1 \\
    0 & 0
    \end{pmatrix}
\end{align}
which is the all-ones matrix between $K^{(1)}$ and $K^{(2)}_x$ (we cannot identify $K^{(1)}$ with $K^{(1)}_x$ and $K^{(3)}_x$ since that would be identifying a single clique with two distinct cliques in the same unit cell). 
The field trace over the conjugate monomial $\bar{x}$ identifies $K^{(2)}$ with $K^{(1)}_{\bar{x}}$, as we expect, and we therefore have that our Krausz decomposition partitions the edges of the graph into all translations of the clique $K = \{1, 1_x, 2_x\}$. 
Note that the vertex $1$ appears in two cliques, $K$ and $K_{\bar{x}}$, and so this vertex corresponds to the edge $(K, K_{\bar{x}})$ in the root graph. 
The vertex $2$ appears in the cliques $K_{\bar{x}}$ and the empty clique $K^{(3)}_{\bar{x}} = \{2\}$, and so it corresponds to an edge $(K_{\bar{x}}, K^{(3)}_{\bar{x}})$. 
Therefore the adjacency matrix of the root graph is given by
\begin{align}
    \compactmat{A}^{\triangle-\mathrm{path \ root}} = 
    \begin{pmatrix}
    K & K^{(3)}
    \end{pmatrix}
    \begin{pmatrix}
    x + \bar{x} & 1 \\
    1 & 0
    \end{pmatrix}
    \begin{pmatrix}
    K \\ 
    K^{(3)}
    \end{pmatrix}
    \label{eq:combmatrix}
\end{align}
which gives the correct root graph.

\section{Compact Laurent-Polynomial Description of Spin Hamiltonians}\label{sec:laurentspinmodels}

Here we will show that the method given above to use binary-valued linear algebra and Laurent polynomials to formulate both compact and non-compact descriptions of graph adjacency matrices can be generalized to encoding bosonizations.
Once the bosonization is formulated in this way, simple linear-algebraic manipulations modulo two can be used to pass from the bosonization to the corresponding frustration graph.

Once the Laurent-polynomial-valued adjacency matrix $\compactmat{A}$ describing a frustration graph is known, our line-graph recognition algorithm, described in Section \ref{sec:lgrecognize}, will determine if the corresponding graph is a line graph, and if so, will also construct the root graph and its compact adjacency matrix. 
Calculation of the corresponding free-fermion band structure can be done by mapping the Laurent monomials $x$ and $\bar{x}$ to the phase difference between unit cells in a Bloch-wave solution, as described in Section \ref{sec:numerics}.
It is therefore desirable to have a description of the spin model itself which is also in compact Laurent-polynomial-valued form and from which frustration graphs can readily be computed. 
To construct this, we will start by encoding the spin model in a non-compact matrix $\fullmat{B}$ and then show that this encoding can be compactified.

In order to formulate the encoding, we represent Pauli operators by $3$-tuples $(b_1, b_2, b_3)$ in the following way: $011$ for $\pauliX$, $101$ for $\pauliY$, $110$ for $\pauliZ$, and $000$ for the identity. 
This encoding is related to our previous notation of using two bits, $(j_x, j_z)$, to specify single-qubit Paulis by the following transformation
\begin{align}
    \begin{pmatrix}
    b_1 \\
    b_2 \\
    b_3
    \end{pmatrix} &=
    \begin{pmatrix}
    0 & 1 \\
    1 & 1 \\
    1 & 0
    \end{pmatrix}
    \begin{pmatrix}
    j_x \\
    j_z
    \end{pmatrix} \mod 2 \mathrm{.}
\end{align}
In general, Hamiltonian terms are written as products of only the relevant non-identity Paulis, but by including the identity explicitly, single-qubit operators can be tensored together such that each Hamiltonian term in a model with $N$ total qubits is a product of $N$ single-qubit operators. 
By concatenating the corresponding $3$-tuples on all qubits, each term in the Hamiltonian can be encoded in a binary vector of length $3\times N$. 
For example, the case of two qubits with a $ZZ$ interaction is described by the vector $110 \, 110$. 
We define a bosonization of the entire Hamiltonian by a binary matrix, $\fullmat{B} \in \mathds{F}_2^{|E| \times 3N}$, where each row represents a single Hamiltonian term:
\begin{align}
    \fullmat{B}_{i, \boldsymbol{q}(j)} &\equiv
    \begin{pmatrix}
    b_{i, 3j - 2} & b_{i, 3j - 1} & b_{i, 3j}
    \end{pmatrix} \nonumber \\
    &= \begin{cases}
    \mathbf{1}_{1 \times 3} - \mathbf{e}_{k} & \mathrm{Pauli} \ k \ \mathrm{is \ at \ qubit} \ j \ \mathrm{in \ term} \ i \\
    \mathbf{0}_{1 \times 3} & \mathrm{otherwise}
    \end{cases} \label{eq:encodingBdef}
\end{align}

We define the matrix $\mathbf{E}$ as 
\begin{align}
 \fullmat{E}_{i, \boldsymbol{q}(j)} &\equiv
    \begin{pmatrix}
    \mathrm{E}_{i, 3j - 2} & \mathrm{E}_{i, 3j - 1} & \mathrm{E}_{i, 3j}
    \end{pmatrix} \nonumber \\
    &= \begin{cases}
    \mathbf{1}_{1 \times 3} & i = j \\
    \mathbf{0}_{1 \times 3} & \mathrm{otherwise}
    \end{cases} \mathrm{,}
\end{align}
such that the kernel of $\mathbf{E}$ is the set of all valid Pauli encodings by this method. The matrix $\mathbf{B}$ satisfies 
\begin{align}
    \fullmat{B} \fullmat{E}^{\mathrm{T}} = \mathbf{0} \mod 2,
\end{align}
and therefore always specifies a valid encoding.

It remains to be shown that a matrix encoding of this form is useful. 
For simplicity consider a Hamiltonian with only two terms encoded in this way. 
The value of their commutator can be found by computing the dot product of the corresponding vectors modulo two. 
To see this, let us compute the dot product modulo two by first taking the dot product mod two within each $3$-tuple. 
If the corresponding Paulis on that qubit anticommute, the result will be one. 
Otherwise it will be zero. 
To complete the calculation, we sum up the results from each qubit-tuple, thereby counting the total number of qubits which contribute an anticommutation. 
Taking this number modulo two completes the calculation and determines whether the total number of minus signs contributed by the constituent single-qubit terms is even or odd. 
As a result, all anticommutation relations between Hamiltonian terms can be found by computing all such dot products, and this non-compact bosonization matrix $\fullmat{B}$ can be used to compute the non-compact adjacency matrix of the frustration graph. 
In fact, when matrix multiplication is taken modulo two, $\fullmat{B}$ and $\fullmat{B}^T$ factorize $\fullmat{A}$:
\begin{align}
    \fullmat{A} = \fullmat{B} \fullmat{B}^{\mathrm{T}} \ \mod 2 \,.
    \label{eq:bosonization}
\end{align}

The Pauli symmetries of the model which can be generated by Hamiltonian terms correspond exactly to binary vectors $\mathbf{v} \in \mathds{F}_2^{\times |E|}$ in the binary kernel of $\fullmat{A}$. 
However, the structure of the symmetry group of the model is defined by the decomposition Eq.~(\ref{eq:bosonization}). 
Specifically, we have
\begin{align}
    \mathrm{ker}(\fullmat{A}) = \mathrm{ker}(\fullmat{B}^{\mathrm{T}}) \cup (\mathrm{im}(\fullmat{B}^{\mathrm{T}}) \cap \mathrm{ker}(\mathrm{\fullmat{B}}))
\end{align}
Operators in $\mathrm{ker}(\fullmat{B}^{\mathrm{T}})$ are products of Hamiltonian terms which give the identity, and the remaining elements of the kernel of $\fullmat{A}$ are products of Hamiltonian terms which multiply to a Pauli that commutes with the Hamiltonian. 
Operators which cannot be made as products of Hamiltonian terms, but nevertheless commute with every operator in the Hamiltonian, correspond to logical-qubit operators. 
These operators correspond to elements of $\left(\mathrm{ker}(\fullmat{B}) \cap \mathrm{ker}(\fullmat{E})\right)/\mathrm{im}(\fullmat{B}^{\mathrm{T}})$.
This gives a rigorous definition of logical operators within this encoding formalism.

It remains to generalize this to a compact formulation of the case where the spin model is translation invariant. 
In this case, we divide the full spin model into unit cells of $M$ qubits such that if any term appears in the Hamiltonian, then translating that term to the corresponding qubits in any other unit cell is also a Hamiltonian term. 
(Note that there will generally be some terms that straddle the boundaries between qubit unit cells.) 
All qubits in the model will then belong to one the of $M$ equivalence classes of qubits, and the individual members of these classes can be indexed by the location of their unit cell in the graph. 
We will encode this location with Laurent polynomials.
Below we present the encoding procedure for the case of two dimensions. The generalization to higher, or lower, dimensions is straightforward.

Each translation-invariant class of Hamiltonian terms can be represented by a single length-$3M$ vector of polynomial-valued $3$-tuples. 
In this case it is simpler to consider only the qubits that contribute non-identity operators to the Hamiltonian term. 
To generate the corresponding compact vector description of this class of Hamiltonian terms, consider one element of the class, and assume that the first participating qubit is in the unit cell at the origin. 
We will start from the zero vector and construct a polynomial-valued vector from the locations of the constituent qubits and the $3$-tuples corresponding to each type of Pauli. 
For each participating physical qubit, we determine its equivalence class $\alpha$, the location of the unit cell that it is in $(n, n')$, and the corresponding Pauli: $\pauliX$, $\pauliY$, or $\pauliZ$. 
This operator will then be encoded as $x^n y^{n'}$ times the $3$-tuple corresponding to that Pauli at position $\alpha$ in the encoding vector. 
If multiple qubits of this class participate the the Hamiltonian term, then we will sum the corresponding tuples. 
This defines the compact Laurent-polynomial-valued encoding matrix $\compactmat{B} \in \mathds{F}_2^{\times t \times 3M}[x, \bar{x}, y, \bar{y}]/\langle x\bar{x} - 1, y\bar{y} - 1 \rangle$, where $t$ is the number of distinct classes of Hamiltonian terms. The constituent block matrices of $\compactmat{B}$ are given by:
\begin{align}
    \label{eq:compactB}
    &   \left(\mathbf{B}_{x^n y^{n'}}\right)_{i, \boldsymbol{q}(j)} \\ 
    & \equiv \begin{cases}
    \mathbf{1}_{1 \times 3} - \mathbf{e}_{k} & \text{qubit $j$ of cell $(n, n')$ in term $i$ is Pauli $k$} \\
    \mathbf{0}_{1 \times 3} & \text{otherwise} 
    \end{cases} \nonumber 
\end{align}
To illustrate this encoding consider a simple one-dimensional example with two spins per unit cell. 
If there is a $YY$ interaction between the two qubits in each unit cell, then this class of Hamiltonian terms is encoded by the vector $101\, 101$. 
A $ZZ$ term between the second qubit in one unit cell and the first qubit in the next unit cell over is given by $xx0 \, 110$. 
The compact encoding matrix is then 
\begin{equation}\
\compactmat{B} = 
\begin{bmatrix}
    101 & 101 \\
    xx0 & 110
\end{bmatrix}.
\end{equation}

Next consider how to find the anticommutation relations between terms encoded in this way. 
Once again, the dot product modulo two of two terms encodes the quantity of interest, but this quantity is now a Laurent polynomial with binary coefficients. 
Consider the dot product between two such term-class encodings. 
If the degree zero term is non-zero, then that indicates that elements of these two classes anticommute when centered on the same unit cell. 

What about the high-order terms? Consider for example a pair of terms which produce an $x$ term in the dot product mod two. 
One way to obtain this term is to have $011$ (i.e.~$\pauliX$) on the first qubit in the unit cell in one term class, and $x0x$ (i.e.~$\pauliY$) in the other term class. 
However, this corresponds to $\pauliY$ on a qubit of the first equivalence class in a unit cell which is one lattice translation to the right.
As a result, when terms from each of these two classes are centered on the origin unit cell, the second term actually involves a physical qubit in the first unit cell to the right. 
This is a distinct physical qubit from the first qubit in the original unit cell, and there is no anticommutation from this pair of terms. 
However, if you consider a pair of terms from these two classes where the first is centered at the original unit cell, and the second is centered at the cell one to the left (i.e.~translated by $\bar{x}$), then this pair of terms acts on the same physical qubit. 
Hence $(011) \cdot (x0x) \mod 2 = x$ indicates that the elements of the two classes of terms anticommute with each other when one is displaced by $x$, or $\bar{x}$, depending on which term the translation is applied to.

This is precisely the information that the compact-valued adjacency matrix of the frustration graph encodes. 
Hence Eq.~(\ref{eq:compactB}) gives a generalization of the incidence matrix to a Bosonization and multiplying it with its transpose conjugate modulo two yields the compact Laurent-polynomial-valued adjacency matrix of the frustration graph: 
\begin{align}
    \compactmat{A} = \compactmat{B} \compactmat{B}^{\dagger} \ \mod 2 \,.
    \label{eq:compactbosonization}
\end{align}

\section{Skew Energy}\label{sec:skewEconnections}

Next we will consider which properties of the root graph spectrum determine the energy gap above the ground state of the spin model.
Under the free-fermion solution to a given spin Hamiltonian, we have
\begin{align}\label{eqn:HDef}
    H_s \mapsto i \sum_{\tau} \left(\boldsymbol{\Gamma} \cdot \mathbf{h}^{(\tau)} \cdot \boldsymbol{\Gamma}^{\mathrm{T}}\right) \Pi_{\tau}\mathrm{.}
\end{align}
Here $\tau$ labels a mutual eigenspace of the symmetries of $H_s$, as specified by a representative orientation of the root graph $R$, and the $\Pi_{\tau}$ are orthogonal projectors onto the respective $\tau$ subspaces. 
In a given symmetry sector, the Hamiltonian can be brought to the diagonal form in Eq.~(\ref{eq:diagonalferm}). 
Assuming that the $(|V| \ \mathrm{mod} \ 2)$ fermionic-parity sector is physical in the spin model, 
the ground state energy of the Hamiltonian in this sector is given by
\begin{align}\label{eq:skewdef}
    \minskew{\tau} = - \sum_{j = 1}^{\lfloor |V|/2 \rfloor} \lambda_{j}^{(\tau)} = - \frac{1}{2}\,  \mathrm{Tr}\left(|\mathbf{h}^{(\tau)}|\right)
\end{align}
When $|h_{j, k}^{(\tau)}| \in {0, 1}$ for all $j$, $k \in V$, then $-2 \minskew{\tau}$ is a graph-theoretic quantity knows as the \emph{skew energy} of the graph $R$ with orientation $\tau$~\cite{li2013survey}. 
Let
\begin{align}
    \begin{cases}
    \tau_1 \equiv \mathrm{argmin}_{\tau} \minskew{\tau}  \\
    \tau_2 \equiv \mathrm{argmin}_{\mathcal{T}_1^\perp} \minskew{\tau}   &
    \mathcal{T}_1^\perp \equiv \{ \tau |  \minskew{\tau}  > \minskew{\tau_1}\}\\
    \end{cases}
\end{align}
be the orientations of the root graph with the two lowest ground state energies, with $\minskew{\tau_1}$  the ground state energy of the original spin model. 

The next-lowest energy of the spin model can either come from a single-particle excitation in the ground-state sector $\tau_1$, governed by the median eigenvalue $\lambda_1^{(\tau_1)}$, or it can come from the ground state energy of a distinct sector $\tau_2$.
Therefore, the gap of the spin model is given by
\begin{align}\label{eqn:relevantGap}
    \Delta = \mathrm{min}\{2\lambda^{(\tau_1)}_1, \minskew{\tau_2} - \minskew{\tau_1}\} \mathrm{.}
\end{align}
Now let us examine the orientation with the lowest ground state energy $\tau_1$. 
For completeness, we relate a given orientation $\tau$ of our root graph to the free-fermion model $\mathbf{h}^{(\tau)}$ by defining the \textit{skew-energy-maximizing orientation}. 
The following theorem was proven in Ref.~\cite{denglan2013skew}. 

\begin{theorem}[Skew-energy-maximizing orientation~{\cite[Thm.~3.11]{denglan2013skew}}]\label{thm:skewmincondition} 
If $R$ has an orientation $\tau$ such that every even cycle is oddly oriented and $|h_{ij}^{(\tau)}|\in \{0, 1\}$ for all $i$, $j \in V$, then $\mathbf{h}^{(\tau)}$  has
the maximal skew energy among all orientations of R.
\label{thm:skewmax}
\end{theorem}

Now let us return to the phase on Eq.~(\ref{eq:cyclephase}) for an even cycle of length $|C| = 2\ell$
\begin{align}
(-1)^c \equiv (-1)^{\tau(C) + \ell}\mathrm{.}
\end{align}
This gives the simple relation between a cycle-symmetry eigenvalue $(-1)^c$, and the orientation of the cycle $\tau(C)$.
It is important to note that the cycle symmetry operator is a \emph{signed} Pauli in general, since it is a Hermitian product of Paulis. 
We will not worry about this detail here. 
What is important, however, is that when we multiply cycle-symmetry operators, associated to say, the cycles $C_1$ and $C_2$, with lengths $2\ell_1$ and $2\ell_2$ respectively, the corresponding eigenvalues $(-1)^{c_1}$ and $(-1)^{c_2}$ multiply. 
The lengths $\ell_1$ and $\ell_2$ do not straightforwardly add however, but the length of the new composite cycle $C_1 \oplus C_2$ rather depends on the geometry of the lattice. 
Thus the orientation of this composite cycle is given by 
\begin{align}
    \tau(C_1 \oplus C_2) = c_1 + c_2 + \ell_1 + \ell_2 - 2|C_1 \cap C_2| \ (\mathrm{mod} \ 2)
\end{align}
This therefore allows us to restate Theorem~\ref{thm:skewmax} in an equivalent, but more physical way.

\begin{theorem}[Skew-energy-maximizing cycle configuration] 
If the $(|V| \ \mathrm{mod} \ 2)$ fermionic-parity sector is physical in the spin model, and there exists a configuration of cycle-symmetry eigenvalues such that
\begin{itemize}
    \item cycles of even, but not doubly even, length have eigenvalue +1, and
    \item cycles of doubly even length have eigenvalue -1,
\end{itemize}
then the global ground state of the spin model is in that configuration.
\end{theorem}
\noindent Since cycle-symmetry eigenvalues multiply to the eigenvalues of their composite cycles, it is therefore sufficient to find a satisfying eigenvalue assignment for an independent generating set of the cycle symmetries and check that the remaining cycle symmetries also satisfy the condition. 
Furthermore, if the root graph is translation invariant, then it is clear that the satisfying eigenvalue configuration, when it exists, is also translation invariant, though the corresponding orientation of the root graph need not be.

Finally, if the $(|V| \ \mathrm{mod} \ 2)$ fermionic-parity sector is not physical, then it is $\minskew{\tau_1} + 2\lambda^{(\tau_1)}_ 1$ which is the lowest energy of the spin model in the sector labeled by $\tau_1$, rather than $\minskew{\tau_1}$. 
We therefore need to check that $\minskew{\tau_1} + 2\lambda^{(\tau_1)}_1$ is not larger than $\minskew{\tau_2} + 2\lambda^{(\tau_2)}_{1}$ in this case. 
Otherwise, it is actually the orientation given by $\tau_2$ which minimizes the energy of the spin model. 
We note here that the graph-theoretic results hold for finite graphs. 
In the infinite translation-invariant case, the relevant corresponding quantity to consider is the energy per particle.
By contrast, the correction $2\lambda_1^{(\tau)}$ is the correction due to a single-particle excitation.

\section{Heuristics and Numerical Study of Example Lattices}\label{sec:numerics}

Unfortunately, the checkerboard-lattice code presented in Section \ref{subsec:doublyfreefermion} corresponds to a free-fermion model on the square lattice, which is gapless, and we know of no free-fermion solvable model with a two-dimensional frustration graph and exact logical qubits which is gapped. 
We have therefore conducted extensive numerical studies combining previous graph-theoretic results~\cite{kollar2020gap} with free-fermion solvable spin models in order to search for examples of spin models with favorable energetic properties as Hamiltonian systems (e.g.~a spectral gap), and with favorable coding properties as subsystem codes (e.g.~exact Pauli logical operators and large distance).
As described in Section \ref{sec:skewEconnections}, the relevant gap of a free-fermionizable spin model is related to two different gaps of the free-fermion solution: one, the energy gap to excite a single fermion from the zero-temperature half-filling state within the ground-state symmetry sector, denoted by $2 \lambda_1^{(\tau_1)}$; and two, the energy difference between the two lowest sectors with distinct total energies: $\minskew{\tau_2} - \minskew{\tau_1}$. 
We will refer to these as the single-particle and sector-energy gaps, respectively. 
The smaller of these two will dictate the properties of the subsystem code.

In order to gain understanding of the behavior of these two gaps, we undertook a study of a series of examples of free-fermion models which contain the solution to a set of exactly solvable spin models with coefficients all zero or one. 
In order to make use of previous graph theoretic results~\cite{kollar2020gap, denglan2013skew}, we restrict our numerical simulations primarily to root graphs which are $3$-regular as well as translation invariant and whose corresponding spin-models have the anticommutation relations of $4$-regular line graphs. 
We generate the root graphs directly using two methods: first, applying one-dimensional periodic boundary conditions on graphene to produce nanotubes, and second using the method of Abelian covers to stitch copies of a finite graph together to form a lattice. 
The latter is a standard mathematical framework for viewing and constructing periodic systems. 
A sketch of the construction and a discussion of its utility will be given in Section \ref{subsec:latticeconstruction}. 
See Ref.~\cite{kollar2020gap, BiggsGraphTheory} for a more detailed discussion. 

The Bloch-wave calculations used to compute the free-fermion single-particle energies and the gaps for different lattices are described in Section~\ref{subsec:numericalresults}. Within this picture, the relevant free-fermion Hamiltonian is equivalent to that of electrons moving in a magnetic field which gives rise to Peierls phases of $\pm i$ on every bond, the sign of which is given by the orientation $\tau$. Cycle orientations and stabilizer eigenvalues therefore map to fluxes or Aharonov-Bohm phases.

\subsection{Lattice Construction}\label{subsec:latticeconstruction}
A numerical search of all possible lattices is not possible, so attention was focused on canonical lattices such as graphene and the square lattice and two sets of examples known from Ref.~\cite{kollar2020gap} to give rise to large $\lambda_1^{(\tau)}$: first, one and two-dimensional square-lattice Abelian covers of small $3$-regular graphs, and second, carbon nanotubes of small diameter.

The Abelian-cover lattices were generated by chaining together copies of a $3$-regular base graph. 
The spirit of this method is very closely related to the Laurent-polynomial-valued compact formulation of the lattice adjacency matrix described above, and it can be thought of as a method for producing an infinite periodic graph from a finite graph $\mathfrak{B}$ by assigning Laurent-polynomial weights to each edge in the base graph. 
Each unit cell of the lattice consists of a copy of all the vertices of $\mathfrak{B}$, and the inter- and intra-unit-cell edges are specified by the Laurent-polynomials chosen in the weighting of the edges of $\mathfrak{B}$. 
If two vertices in the base graph are connected by an edge of weight $1$, then their images are connected within each copy of the unit cell. 
If the edge $(v_1,v_2)$ has weight $x$, then this corresponds to a bond in the lattice between $v_1$ in one unit cell and $v_2$ in the unit cell neighboring it in the positive $x$ direction, similarly for $\bar{x}, y, \bar{y}$, etc.

Any lattice can be constructed in this way starting from either a graph or a multigraph $\mathfrak{B}$, and this method provides three useful features. 
First, the $k=0$ eigenvalues of the oriented lattice are equal to those of the oriented base graph. 
This allows some rudimentary filtration for unit cells and orientations which have a chance to give rise to lattices with large gaps at zero energy. 
Second, the possible small regular graphs have been tabulated (see e.g.~Ref.~\cite{CDS} for the $3$-regular case). 
Third, and most significantly for this formalism, it provides a convenient framework for systematically reducing the number of orientations of the root graph which must be computed in order to estimate the sector-energy gap $\minskew{\tau_2}- \minskew{\tau_1}$. 
For an orientation which repeats every $n$ unit cells, the corresponding free-fermion model on the root graph has a magnetic unit cell which consists of at most $n$ unit cells. 
If there are $m$ bonds in the base graph, then naively there can be as many as $2^{m\times n}$ orientations which are periodic on this length scale. 
Computing a band structure and density of states for each of these orientations would generally be prohibitively expensive. 
Fortunately, many of these orientations are redundant and correspond to the same sets of fluxes, or even no flux at all. 
The possible distinct flux configurations can be enumerated by noting that the magnetic model corresponding to each orientation is naturally constructed as an Abelian cover of an enlarged base graph $\mathfrak{B}^{(n)}$. 
While this construction is geometrically redundant, the flux through any closed loop of the full lattice is uniquely determined by the Peierls phases on the edges of $\mathfrak{B}^{(n)}$. 
The possible flux configurations of $\mathfrak{B}^{(n)}$ can be found by choosing a spanning tree and varying the orientation of the edges not in the spanning tree, which vastly reduces the number of options that need to be computed. 
For highly symmetric base graphs, such as the cube, the true number of configurations is even lower than this due to further symmetries~\cite{chapman2020characterization, yaoping2011oriented}.

Using the tabulation in Ref.~\cite{CDS}, a series of Abelian cover lattices were constructed from select $3$-regular base graphs with 12 or fewer vertices. 
To restrict the search space, we consider only covers where the covering group is $\mathbb{Z}$ or $\mathbb{Z} \times \mathbb{Z}$, corresponding to one-dimensional chains, and two-dimensional square lattices. 
Furthermore, we restrict to covers where there are at most two edges that produce inter-unit-cell connections. 
In the Laurent-polynomial formulation, this corresponds to giving all but two edges the monomial $1$. 
The remaining two edges may either be $\{ 1, x\}$, $\{x,y\}$, $\{x,x\}$, or $\{x,\bar{x}\}$. 
This is a severely restricted set of covers; however, it was shown in previous work~\cite{kollar2020gap} that it contains examples which exhibit the largest possible gaps for symmetric adjacency matrices of \emph{arbitrary} unoriented 3-regular graphs. 
The numerical studies described below show that it also contains examples with extremely large values of $\lambda_1^{(\tau)}$.
In fact, the $(2,0)$-nanotube shown in Fig.~\ref{fig:hourglass} is one such example with $2\lambda_1^{(\tau)} = 2$~\cite{kollar2020gap,MoharGap,MoharMedian}.
The $(1,1)$-nanotube, or equivalently the ladder, shown in Fig.~\ref{fig:ladder}, is another example with $2\lambda_1^{(\tau)} = 2$. However, in this case, this large single-particle gap is found in the absolute ground state orientation.
Interestingly, both of these apparently extremal examples arise as $\{x,x\}$ Abelian covers of the cube.

\begin{figure}[t!]
\centering
		\includegraphics[width=0.45\textwidth]{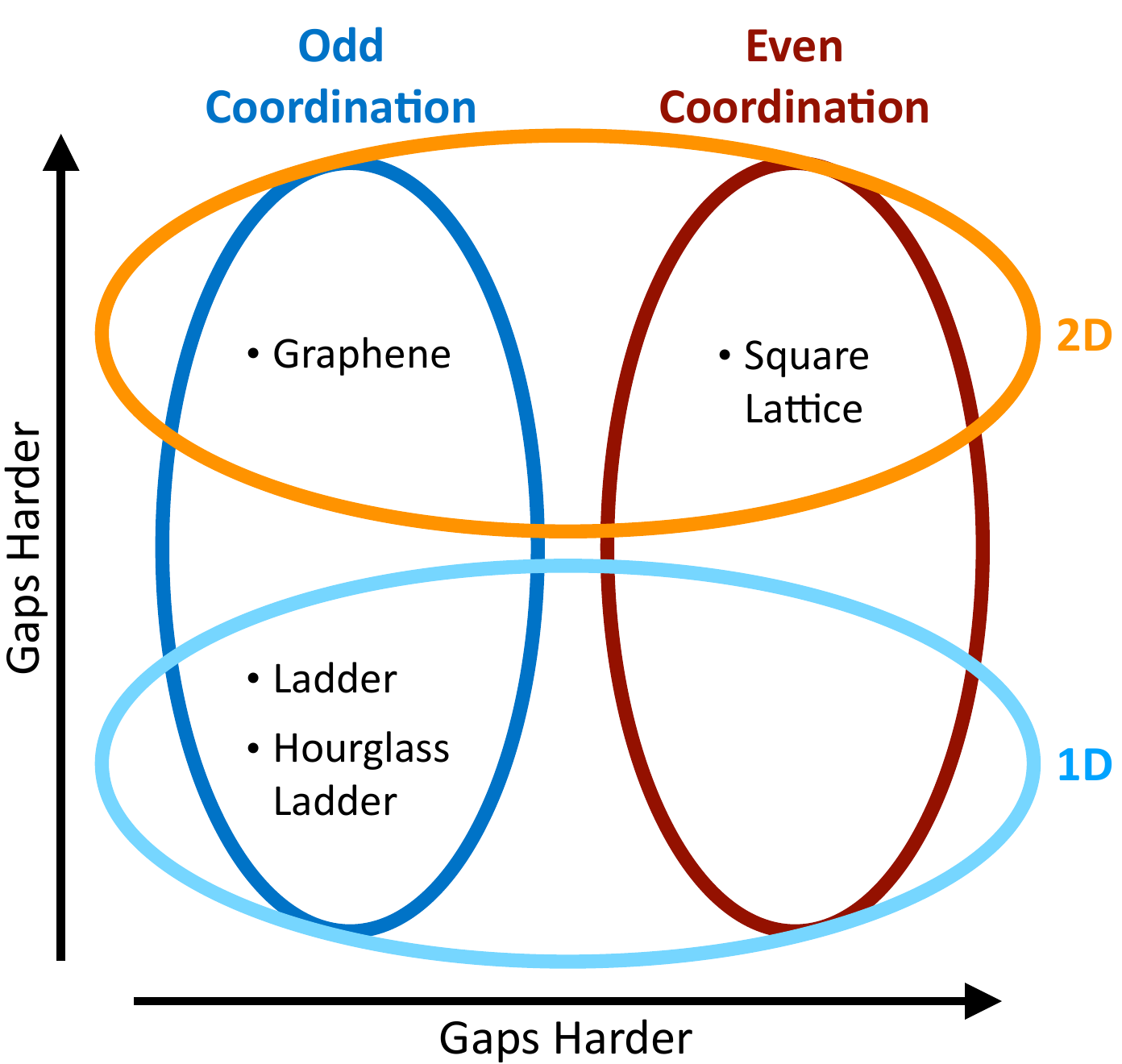}
	\caption{Schematic diagram indicating which classes of lattices tend to exhibit large single-particle gaps at half filling and where certain special examples fall. 
	In general, higher-dimensional lattices tend to have fewer and smaller gaps due to their larger phase spaces. 
	Lattices with odd coordination numbers often have very low densities of states near zero energy, making it much easier to find variations with large gaps or to introduce gaps by perturbations to the hopping coefficients. 
	As a result, it is very difficult to open up a gap at zero energy for cases like the square lattice (2D, and even-coordinated), and significantly easier to do so for graphene (2D, odd-coordinated), whereas the largest single-particle gaps are found in cases like the hourglass ladder (see Fig.~\ref{fig:hourglass}) and the ladder (see Fig.~\ref{fig:ladder}),  which are quasi-one-dimensional and odd-coordinated.}
	\label{fig:ven}
\end{figure}

\begin{figure*}[t!]
\centering
		\includegraphics[width=0.70\textwidth]{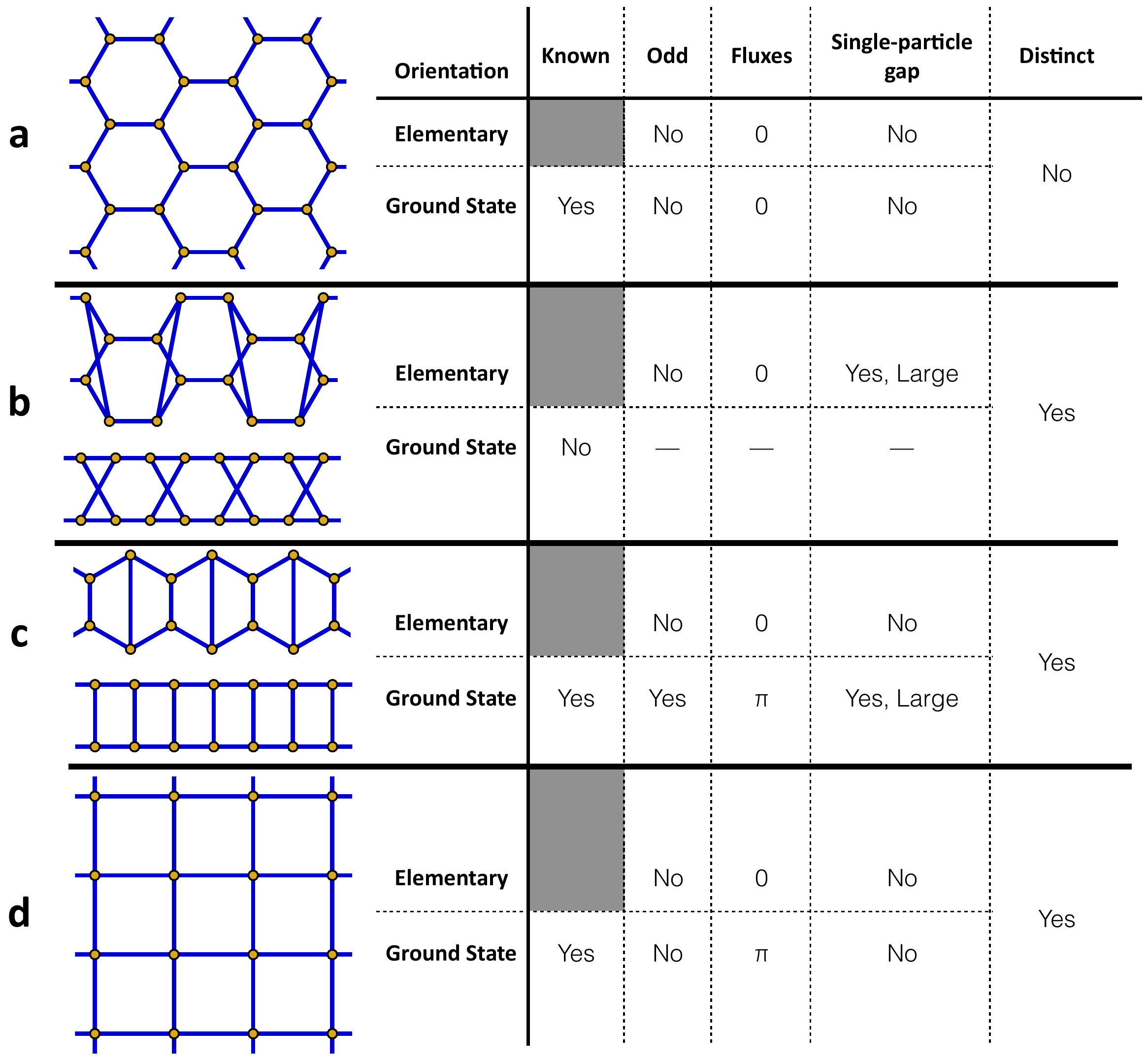}
	\caption{Table of known properties for the four main examples in Fig.~\ref{fig:ven}: \textbf{a} graphene, \textbf{b} the $n=2$, $m=0$ carbon nanotube, \textbf{c} the $n=1$, $m=1$ carbon nanotube, and \textbf{d} the square lattice. 
	For each graph, the cycle-orientation, plaquette fluxes, and single particle gap are tabulated for two key orientations: 
	 the elementary orientation and the ground state (if known). 
	 In the case of graphene, these two orientations are the same, whereas in the three others they are distinct. 
	 Both nanotubes have simpler realizations as a ladder or hourglass ladder, and both equivalent versions of the graph are shown. 
	 The two orientations of \textbf{b} and \textbf{c} and the accompanying densities of states are shown in Fig.s~\ref{fig:hourglass} and \ref{fig:ladder}, respectively. 
	 Each example embodies the trends shown in Fig.~\ref{fig:ven} for creation of single-particle gaps, with the quasi-one-dimensional and $3$-coordinated nanotubes exhibiting the largest single-particle gaps. }
	\label{fig:quad}
\end{figure*}

\subsection{Gap Calculations}\label{subsec:numericalresults}

For each translation-invariant root graph, we use Bloch-wave theory to compute the single-particle band structure of the root graph with Peierls phases of $\pm i$ on each bond and with all hopping amplitudes equal to one. 
These phases encode the orientation $\tau$ of the graph, and the Hermitian single-particle Hamiltonian of this model is equal to $i$ times the antisymmetric adjacency matrix $\bf{h}^{(\tau)}$, as shown in Eq.~\ref{eqn:HDef}, and the Hamiltonian for a given $\vec{k}$ is $i \compactmat{A}$ with $\boldsymbol{x}_j \rightarrow \exp{i \vec{k} \cdot \vec{a_j}}$, where $\vec{a_j}$ is the $j^{th}$ lattice vector. 
We numerically compute the density of states, as well as the total energy and single-particle gap for each distinct orientation below a maximum magnetic unit cell size, typically up to a few times the geometric unit cell size. 
In general, there is no mathematical guarantee that the ground state symmetry sector must be of the form above. 

However, for uniform hoppings, the range in which eigenvalues can exist is dictated by the coordination number, and the bands become more and more tightly packed as the unit cell size becomes large, making the chances of finding a ground state symmetry sector with a large single-particle gap very low. 
Therefore, favorable models which exhibit large gaps in both senses should fall within the scope of the numerical search. 
The outcome of the limited numerical search can thus be viewed as a likely indicator for how a given root graph will perform, and in some cases, where the absolute ground state symmetry sector is known by other means (e.g.~Ref.~\cite{lieb1994flux} or Thm.~\ref{thm:skewmincondition}), stronger statements can be made.

Combining these numerical results with previous graph-theoretic ones examining large gaps in the spectra of regular graphs~\cite{kollar2020gap}, we find a few general principles, which are indicated schematically in Fig.~\ref{fig:ven}. 
A set of examples that highlights these principles is shown in Fig.~\ref{fig:quad}, and we will discuss these in detail below.

The excitation gap $2\lambda_1^{(\tau)}$ is dictated by the single-particle spectrum, and so a fair amount is known about its behavior from both mathematical studies of graph spectra~\cite{kollar2020gap, CDS} and also from the study of band structures in the solid state, e.g.~\cite{GirvinSS}.
In general, large single-particle excitation gaps ($\lambda_1^{(\tau)} = \mathcal{O}(1)$) are increasingly difficult to achieve in higher dimensions due to the larger phase space, and largest values of $\lambda_1^{(\tau)}$ are generally found in quasi-one-dimensional lattices~\cite{kollar2020gap}. 
As a result, there is a fundamental tension between the desire for higher-dimensional lattices which can have much richer topological properties and the desire for large single-particle gaps which give rise to intrinsic suppression of local errors.

Despite some dramatic exceptions which host compact-support states similar to those found in the Lieb lattice~\cite{Kollar:2019linegraph}, root graphs with odd coordination number generally have difficulty sustaining a large number of states in which each site has an equal number of positive and negative neighbors. 
The resulting suppression of the density of states near zero energy makes it easier for an orientation which gives rise to a staggered magnetic field, or a set of non-uniform hoppings, to open up a single-particle gap. 
For example, the square lattice in zero magnetic field has a ring of momenta with zero energy, whereas graphene has exactly two points at zero energy, one from each Dirac cone. 
Correspondingly, graphene is much easier to gap out than the square lattice.
Thus, intuitively, the largest values $\lambda_1^{(\tau)}$ should be found in quasi-one-dimensional root graphs with odd-coordination number and relatively small magnetic unit cell sizes. 

\begin{figure}[th!]
\centering
		\includegraphics[width=0.45\textwidth]{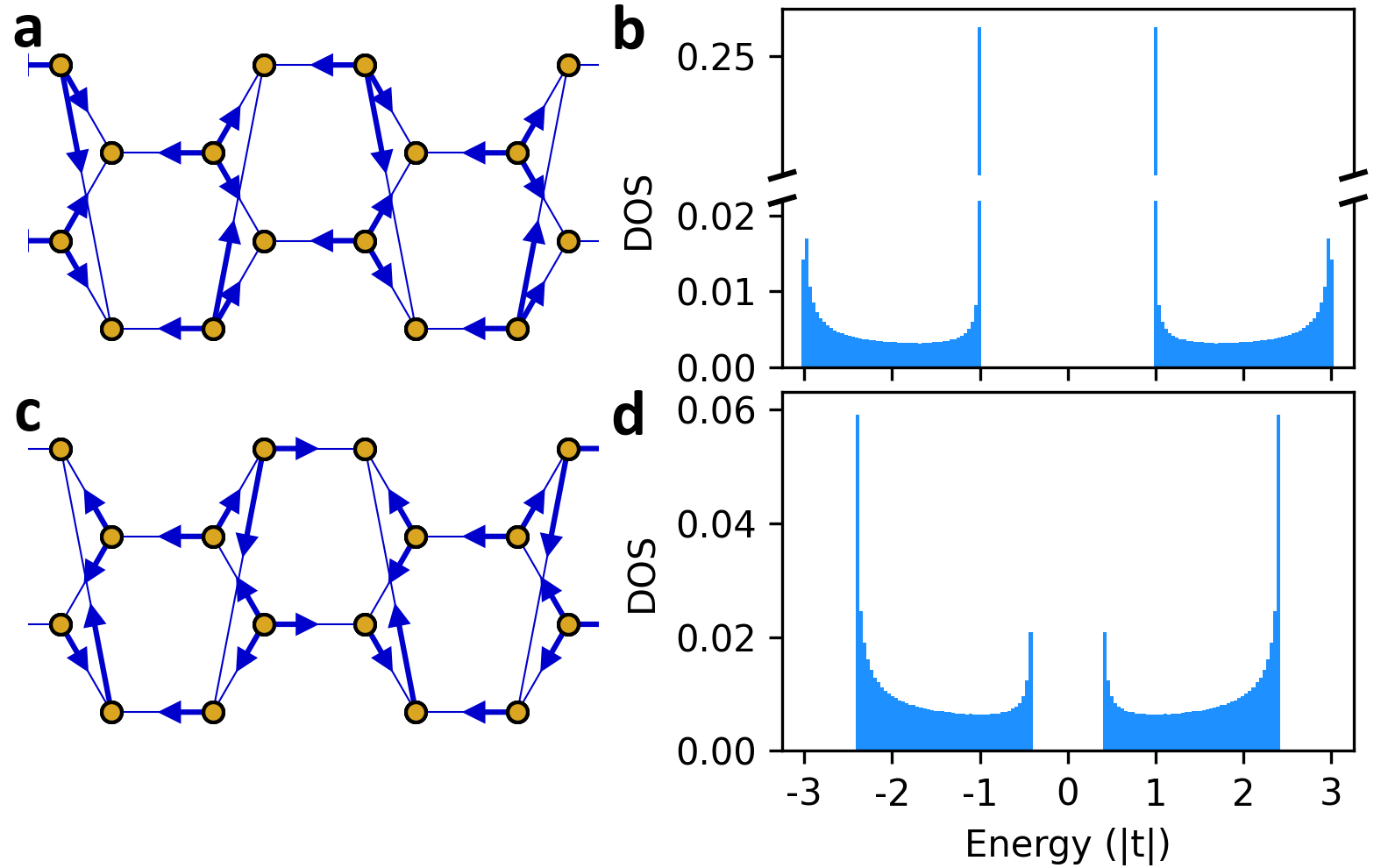}
	\caption{$n = 2$, $m = 0$ nanotube. 
	The unoriented version of this nanotube is one graph which achieves the largest gap possible for any $3$-regular graph~\cite{kollar2020gap,MoharGap}. 
	The elementary orientation is shown in \textbf{a}, with arrows on each edge indicating their orientation. 
	The elementary orientation necessarily has the same DOS (\textbf{b}) as the unoriented graph. 
	Despite the presence of a peak in the DOS at $-3$ and the absence of states in the interval $(-1,0]$, this orientation is not the ground state. 
	\textbf{c} and \textbf{d} show the corresponding plots for a lower-energy orientation. 
	This orientation has no eigenvalues near $-3$ and there are states above $-1$, but it does not have a flat band at $-1$ containing a quarter of all the states. 
	As a result, the total energy per particle at half-filling is $\sim 0.033$ lower than in the elementary orientation. }
	\label{fig:hourglass}
\end{figure}

\begin{figure}[th!]
\centering
		\includegraphics[width=0.45\textwidth]{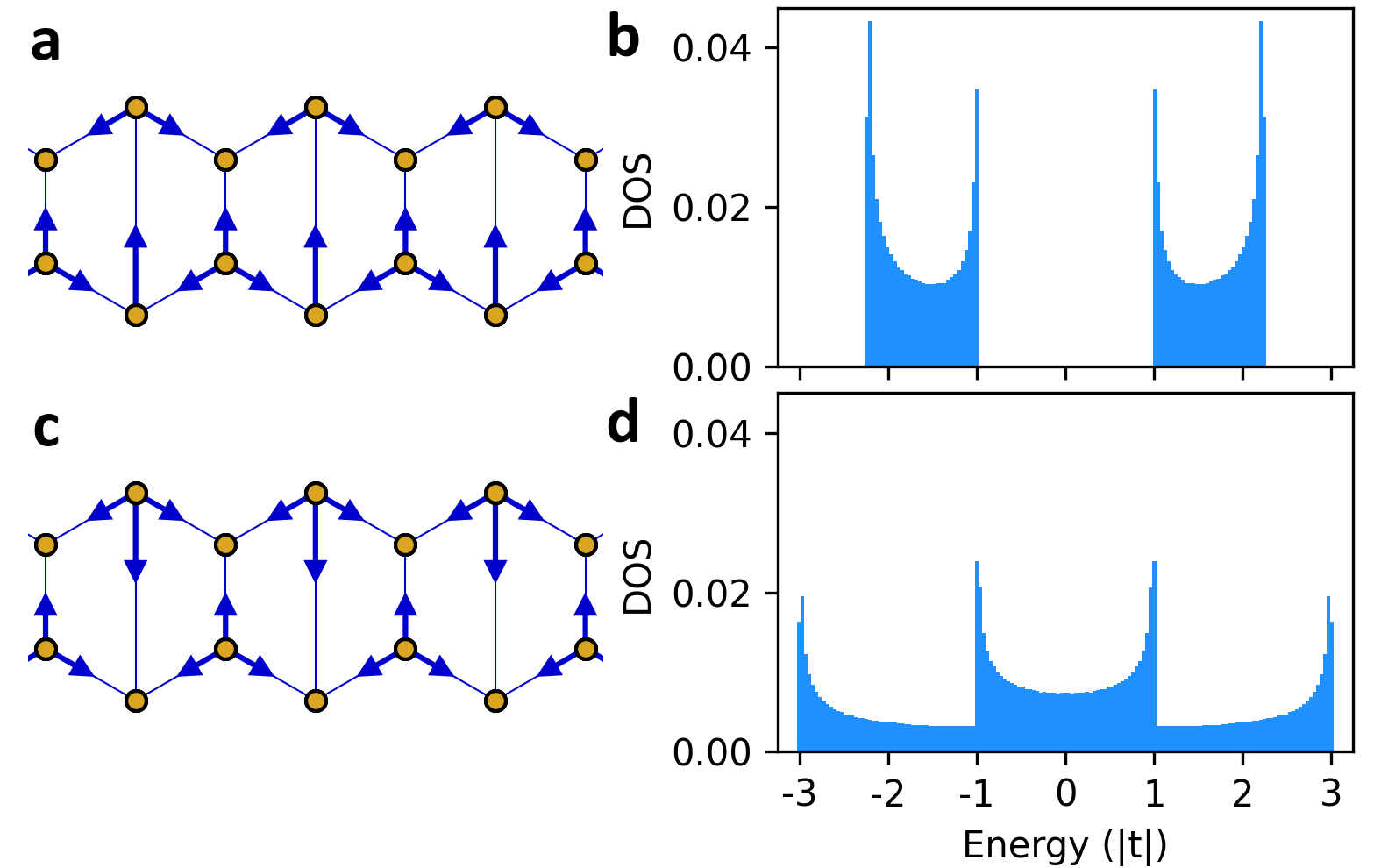}
	\caption{$n = 1$, $m = 1$-nanotube. 
	This nanotube, or ladder, is an example of a graph where the exact ground-state orientation is known.
	Its elementary orientation and the corresponding DOS are shown in \textbf{c} and \textbf{d} and exhibit no single-particle gap. 
	The corresponding plots for the true ground-state orientation are shown in \textbf{a} and \textbf{b}. 
	This orientation has an extremely large single-particle gap: $2 \lambda_1^{(\tau)} = 2$. 
	By contrast, the sector-energy gap is bounded above by $\sim 0.068$ per particle.}
	\label{fig:ladder}
\end{figure}

The numerical results corroborate this intuition. 
The two most dramatic single-particle gaps found are show in Figs.~\ref{fig:hourglass} and \ref{fig:ladder}, each with $2\lambda_1^{(\tau)} = 2$. 
Both arise in a quasi-one-dimensional $3$-regular root graph, with a magnetic unit cell containing $4$ distinct sites. 
On the flip side, for modest magnetic unit-cell sizes, the two dimensional square lattice, whose coordination number is even, never exhibits a single-particle gap.

In contrast to the single-particle gap $2\lambda_1^{(\tau)}$, the total energy $\minskew{\tau}$, which is determined by the sum of the energies of all the occupied states, is a more unusual quantity about which much less is known. 
The corresponding mathematical quantity, the skew energy~\cite{adiga2010skew}, was introduced only recently, and its value for lattice models is not commonly considered in solid-state physics. 
Unlike $\lambda_1^{(\tau)}$, which depends only on the density of states near the Fermi energy at half filling, $\minskew{\tau}$ depends sensitively on the energies of bands far below the Fermi surface. 
A dramatic example of this is found in the case of the $n = 2$, $m = 0$ carbon nanotube shown in Fig.~\ref{fig:hourglass}, which is equivalent to the hourglass ladder discussed in~\cite{kollar2020gap, MoharGap}.
Since this graph is bipartite, it has an elementary orientation, shown in Fig.~\ref{fig:hourglass}\textbf{a}, in which all edges are oriented to point away from one of the two sublattices and toward the other. 
The elementary orientation always corresponds to no magnetic flux, and its energies are thus equivalent to those of the unoriented case. 
In Ref.~\cite{kollar2020gap, MoharGap}, it was shown that the spectrum of this graph has the largest possible gap that an unoriented $3$-regular graph with unit hopping can have. 
However, this large gap is accompanied by completely flat bands at $\pm 1$ which comprise half of the total states. 
These pull the total energy up, and there exist lower-energy symmetry sectors such as that shown in Fig.~\ref{fig:hourglass}\textbf{c}. 

Another dramatic example arises in the case of the $n = 1$, $m = 1$ carbon nanotube shown in Fig.~\ref{fig:ladder}, whose graph is equivalent to the ladder. 
In this case, the elementary orientation, shown in Fig.~\ref{fig:ladder}\textbf{a}, does not exhibit a single-particle gap. 
However, in the case of this graph the true ground-state symmetry sector is known from Thm.~\ref{thm:skewmincondition}~\cite{denglan2013skew}, and it does exhibit a large $\lambda_1^{(\tau)}$, as shown in Fig.~\ref{fig:ladder}\textbf{b}. 
In general, we find very little correlation between large single-particle gaps of the non-magnetic lattice, large $\lambda_1^{(\tau)}$, and minimizing $\minskew{\tau}$.

Furthermore, while the numerical search produced a handful of quasi-one-dimensional and even two-dimensional examples  which have a symmetry sector with $2 \lambda_1^{(\tau)} > 0.8$, no examples were found with $\minskew{\tau_2}- \minskew{\tau_1}$ similarly large. 
Assuming that the true ground-state sector has a small magnetic unit cell, and is therefore among those computed, the true sector-energy gap bounded above by the smallest numerically observed gap, up to numerical error.
Thus, for all computed examples, the sector-energy gap is observed to be small, $\minskew{\tau_2}- \minskew{\tau_1} < 0.1$ per particle. 
It is not clear at this stage whether the relatively small sector-energy gaps are indicative of a fundamental constraint, or simply that we lack an understanding of which graph-level properties are required to produce a well-isolated ground-state manifold with a significant gap to the next-lowest set of orientations. 

Since the total energy of an orientation depends on an integral over the band structure, opening up small gaps by changing the weighting of the edges will generally shift as many states up in energy as it does down, and is therefore unlikely to have significant effects. 
The strongest effects will occur only when band centers shift. 
Furthermore, the sector-energy gap involves comparing different symmetry sectors whose effective magnetic field patterns differ by fluxes $\mathcal{O}(\Phi_0/4)$. 
These enormous fields can alter the band structure radically, making it extremely difficult to predict which band structures are possible or identify the ground-state configuration. 
If the latter is known, it is generally only from combinatorial arguments~\cite{denglan2013skew, lieb1994flux}. 

Additionally, obtaining a reasonable bound on the sector-energy gap $\minskew{\tau_2}- \minskew{\tau_1}$ requires determining not only the ground state, but also searching over larger and larger magnetic unit cells with more and more possible orientations in order to identify the lowest-excited orientation. 
As a result, the sector-energy gap is computationally intensive to study, and it was computed intensively only for graphs for which were already known to exhibit large values of $\lambda_1^{(\tau)}$.

In sum, despite the fact that large single-particle gaps $2\lambda_1^{(\tau)}$ do not occur in the spectrum of a generic graph, the search space of Abelian covers of small $3$-regular graphs (which includes small-diameter nanotubes) considered here includes quite a few such graphs where at least one orientation has a large $\lambda_1^{(\tau)}$. 
The most dramatic such examples are shown in Figs.~\ref{fig:hourglass} and \ref{fig:ladder}. 
In contrast, much less is known about where to find examples with large sector-energy gaps $\minskew{\tau_2}- \minskew{\tau_1}$. 
In all of the examples computed here, none were found in which $\minskew{\tau_2}- \minskew{\tau_1}$ could be greater than $0.1$, indicating that, thus far, the sector-energy gap is the primary limiting factor in potential energetic suppression of local errors.

\section{Discussion}

In this work, we have given the first examples of 2D free-fermion models that are also exact quantum subsystem codes. 
We obtained these examples using a general formalism for describing translation-invariant free-fermion models, and we have provided several tools for recognizing such models and embedding them into spin models with favorable error suppression properties.
Our numerical searches for models with the most favorable energetics indicate that the relevant energy gap which limits error suppression is not the single-particle energy gap in a given fermion sector, but rather it is the energy gap between ground states of two sectors that is the bottleneck. 
There are currently no known methods for identifying graphs for which this sector-energy gap is large, and it is now clear that developing such methods will be critical to finding subsystem codes with strong intrinsic error suppression.

There are numerous open questions suggested by this work. 
The examples we have exhibited in 2D would be interesting to generalize with new codes in higher dimensions or with more exotic geometries.
Such codes can potentially be realized using a similar strategy to that used to produce the checkerboard-lattice code in Sec.~\ref{subsec:doublyfreefermion}, whereby we ``added edges'' to the frustration graph of the 2-d Bacon-Shor code to complete it to a line graph (and thus a free-fermion model). 
This same strategy could in-principle be used to generate new free-fermion codes in other geometries, thereby leveraging our existing wealth of knowledge of subsystem codes to generate free-fermion-solvable error-correcting models.
Such codes may further be considered a starting point for the treatment of non-integrable models in the framework of error correction (see Ref.~\cite{wildeboer2021symmetry} for a recent investigation into this question).

While we have primarily explored the gauge-Hamiltonian energetics for the purposes of error-suppression, there are potentially other applications of the free-fermion solution for error-correction.
In principle, we can understand these codes as a kind of ``tailored'' error correction code for noise that is biased toward high-energy processes.
Given the recent success of tailored error-correcting codes for biased noise~\cite{tuckett2018ultrahigh, tuckett2019tailoring, tuckett2020faulttolerant} It would be interesting to explore the degree to which the free-fermion solution informs decoding.

Finally, we expect that our tools will be useful for the experimental implementation of such codes.
Given an experimental apparatus where non-commuting short-range interactions can be engineered, it may be natural to expect that we can realize induced subgraphs of a given global frustration graph by turning certain interactions off.
Our results therefore provide a target for experimental platforms to realize error-correction models that be exactly analyzed.

\begin{acknowledgements} 
AC acknowledges support from EPSRC under agreement EP/T001062/1, and from EU H2020-FETFLAG-03-2018 under grant agreement no.~820495 (AQTION). 
This work is supported in part by the Australian Research Council (ARC) via the Centre of Excellence in Engineered Quantum Systems (EQuS) project number CE170100009. 
AJK acknowledges support from AFOSR Grant No.~FA95502110129 and NSF Grant No.~PHY2047732.
\end{acknowledgements}

\appendix
\section{Description of Numerical Simulation Packages}

The numerical calculations in the work were carried out using a custom suite of graph-theoretic lattice codes defined in Python3. 
The backbone of this code was developed for Refs.~\cite{Kollar:2019linegraph,kollar2020gap} and provides convenient automation for generating lattice unit cells, identifying the unit cells of their line graphs, and numerically integrating Bloch-wave solutions to compute densities of states. 
This automation, combined with an extension to imaginary hopping coefficients, enabled the numerical search described in Section \ref{sec:numerics}.

A second set of codes was used to define and search for logical degrees of freedom of spin models defined on qubits associated to the edges or vertices of these lattices. 
The checkerboard-lattice code described in Section \ref{subsec:doublyfreefermion} was discovered during numerical simulation of the fiducial bosonization of the line graph of the square lattice using the following iterative pseudo-algorithm.

First, select a target lattice from the code base and generate a fiducial bosonization Hamiltonian $H_0$ in compact Laurent-polynomial form. 
This can be done either automatically, or manually for higher symmetry. 
Second, manually input any known local stabilizers $S_0$. 
Third, select a system size, apply periodic boundary conditions, and generate the exact Pauli operators corresponding to all translates of each type of Hamiltonian and stabilizer term. 
Using the symplectic encoding of single-qubit Pauli operators into pairs of binary numbers described in Eq.~\ref{eqn:steveencoding}, the set of all operators $C_0$ on the torus which commute with both $H_0$ and $S_0$ can be computed using binary-valued linear algebra. 

Fourth we divide the elements of $C_0$ into stabilizers $\tilde{S}$ and monogamously anticommuting qubit pairs $\tilde{X}$ and $\tilde{Z}$. 
To do so, we loop over the elements $c_n \in C_0$. If $c_n$ commutes with all elements of $C_0$, then we assign it to $\tilde{S}$ and remove it from $C_0$. 
If it anticommutes with another element $c_m$, then we assign $c_n$ to $\tilde{X}$ and $c_m$ to $\tilde{Z}$ and remove them both from $C_0$. 
Any remaining elements that anticommute with $c_n$ are multiplied by $c_m$ to produce commuting operators and vice versa. 
At the end of this process $C_0$ is empty and a (possibly modified) version of each original element is assigned to either $\tilde{S}$, or the set of logical qubits $(\tilde{X}, \tilde{Z})$. 
This new set of operators has clean anticommutation relations, but will generally also contain many non-local operators due to the monogamization step. This completes one iteration of the algorithm.

In general, the first iteration yields an extensive number of messy quasi-local logical degrees of freedom which do not reflect the translation-invariant structure of the starting spin model. 
However, visual inspection of the numerical operators often reveals local operators that nearly repeat at different locations on the torus. 
The art, and non-deterministic aspect, of this method lies in intuiting clean local operators from the set of distorted numerical operators. 
Any local logical operators identified are then encoded in Laurent-polynomial form and eliminated by incorporating one half of each logical pair into an updated set of stabilizers $S_1$. 
The algorithm is then repeated to yield a new set of operators $C_1$ which commute with $H_1$ and $S_1$. 
Note that sometimes it may be necessary to double the size of the unit cell to add every other instance of an operator to the set of stabilizers, and that local operators whose translates form a connected frustration graph which is a line graph can also be added as new Hamiltonian terms.

If a model is trivial and contains no topological logical degrees of freedom, this process will terminate when there are no logical degrees of freedom left. 
If a model has exact topological logical degrees of freedom, then a finite number of non-local operators will remain, and the number of such operators will be (largely) independent of system size. 
However, it is important to note that the choice of the the initial stabilizer set $S_0$ and subsequent updates can effect the number of logicals and even whether or not logicals exist. 
For example, a single fiducial bosonization of the line graph of the square lattice admits dimer-like local operators which commute with the free-fermion Hamiltonian, in addition to the loop-like stabilizers used in the checkerboard-lattice code. 
Initiating the algorithm with a commuting subset of these dimers produces two intertwined free-fermion models with the same double square-lattice frustration graph, but no logicals at all.

Finally, the numerical logicals, while often visually string-like, tend to still be distorted by unwanted multiplication with stabilizers and generally meander. 
Identification of clean logical degrees of freedom must again be done my inspection. It can then be verified that these operators are linearly independent of the Hamiltonian and stabilizer terms and that they exist with the appropriate anticommutation relations for all (or all even) system sizes.


\begin{thebibliography}{58}%
\makeatletter
\providecommand \@ifxundefined [1]{%
 \@ifx{#1\undefined}
}%
\providecommand \@ifnum [1]{%
 \ifnum #1\expandafter \@firstoftwo
 \else \expandafter \@secondoftwo
 \fi
}%
\providecommand \@ifx [1]{%
 \ifx #1\expandafter \@firstoftwo
 \else \expandafter \@secondoftwo
 \fi
}%
\providecommand \natexlab [1]{#1}%
\providecommand \enquote  [1]{``#1''}%
\providecommand \bibnamefont  [1]{#1}%
\providecommand \bibfnamefont [1]{#1}%
\providecommand \citenamefont [1]{#1}%
\providecommand \href@noop [0]{\@secondoftwo}%
\providecommand \href [0]{\begingroup \@sanitize@url \@href}%
\providecommand \@href[1]{\@@startlink{#1}\@@href}%
\providecommand \@@href[1]{\endgroup#1\@@endlink}%
\providecommand \@sanitize@url [0]{\catcode `\\12\catcode `\$12\catcode
  `\&12\catcode `\#12\catcode `\^12\catcode `\_12\catcode `\%12\relax}%
\providecommand \@@startlink[1]{}%
\providecommand \@@endlink[0]{}%
\providecommand \url  [0]{\begingroup\@sanitize@url \@url }%
\providecommand \@url [1]{\endgroup\@href {#1}{\urlprefix }}%
\providecommand \urlprefix  [0]{URL }%
\providecommand \Eprint [0]{\href }%
\providecommand \doibase [0]{http://dx.doi.org/}%
\providecommand \selectlanguage [0]{\@gobble}%
\providecommand \bibinfo  [0]{\@secondoftwo}%
\providecommand \bibfield  [0]{\@secondoftwo}%
\providecommand \translation [1]{[#1]}%
\providecommand \BibitemOpen [0]{}%
\providecommand \bibitemStop [0]{}%
\providecommand \bibitemNoStop [0]{.\EOS\space}%
\providecommand \EOS [0]{\spacefactor3000\relax}%
\providecommand \BibitemShut  [1]{\csname bibitem#1\endcsname}%
\let\auto@bib@innerbib\@empty
\bibitem [{\citenamefont {Jordan}\ and\ \citenamefont
  {Wigner}(1928)}]{jordan1928uber}%
  \BibitemOpen
  \bibfield  {author} {\bibinfo {author} {\bibfnamefont {P.}~\bibnamefont
  {Jordan}}\ and\ \bibinfo {author} {\bibfnamefont {E.}~\bibnamefont
  {Wigner}},\ }\href {\doibase 10.1007/BF01331938} {\bibfield  {journal}
  {\bibinfo  {journal} {Zeitschrift f{\"u}r Physik}\ }\textbf {\bibinfo
  {volume} {47}},\ \bibinfo {pages} {631} (\bibinfo {year} {1928})}\BibitemShut
  {NoStop}%
\bibitem [{\citenamefont {Kitaev}(2006)}]{kitaev2006anyons}%
  \BibitemOpen
  \bibfield  {author} {\bibinfo {author} {\bibfnamefont {A.}~\bibnamefont
  {Kitaev}},\ }\href {\doibase https://doi.org/10.1016/j.aop.2005.10.005}
  {\bibfield  {journal} {\bibinfo  {journal} {Annals of Physics}\ }\textbf
  {\bibinfo {volume} {321}},\ \bibinfo {pages} {2 } (\bibinfo {year} {2006})},\
  \bibinfo {note} {january Special Issue}\BibitemShut {NoStop}%
\bibitem [{\citenamefont {Terhal}\ and\ \citenamefont
  {DiVincenzo}(2002)}]{terhal2002classical}%
  \BibitemOpen
  \bibfield  {author} {\bibinfo {author} {\bibfnamefont {B.~M.}\ \bibnamefont
  {Terhal}}\ and\ \bibinfo {author} {\bibfnamefont {D.~P.}\ \bibnamefont
  {DiVincenzo}},\ }\href {\doibase 10.1103/PhysRevA.65.032325} {\bibfield
  {journal} {\bibinfo  {journal} {Phys. Rev. A}\ }\textbf {\bibinfo {volume}
  {65}},\ \bibinfo {pages} {032325} (\bibinfo {year} {2002})}\BibitemShut
  {NoStop}%
\bibitem [{\citenamefont {{Knill}}(2001)}]{knill2001fermionic}%
  \BibitemOpen
  \bibfield  {author} {\bibinfo {author} {\bibfnamefont {E.}~\bibnamefont
  {{Knill}}},\ }\href@noop {} {\bibfield  {journal} {\bibinfo  {journal} {ArXiv
  e-prints}\ } (\bibinfo {year} {2001})},\ \Eprint
  {http://arxiv.org/abs/arXiv:quant-ph/0108033} {arXiv:quant-ph/0108033}
  \BibitemShut {NoStop}%
\bibitem [{\citenamefont {Valiant}(2002)}]{valiant2002quantum}%
  \BibitemOpen
  \bibfield  {author} {\bibinfo {author} {\bibfnamefont {L.~G.}\ \bibnamefont
  {Valiant}},\ }\href {\doibase 10.1137/S0097539700377025} {\bibfield
  {journal} {\bibinfo  {journal} {SIAM Journal on Computing}\ }\textbf
  {\bibinfo {volume} {31}},\ \bibinfo {pages} {1229} (\bibinfo {year}
  {2002})},\ \Eprint
  {http://arxiv.org/abs/https://doi.org/10.1137/S0097539700377025}
  {https://doi.org/10.1137/S0097539700377025} \BibitemShut {NoStop}%
\bibitem [{\citenamefont {Bravyi}\ and\ \citenamefont
  {Kitaev}(2002)}]{bravyi2002fermionic}%
  \BibitemOpen
  \bibfield  {author} {\bibinfo {author} {\bibfnamefont {S.~B.}\ \bibnamefont
  {Bravyi}}\ and\ \bibinfo {author} {\bibfnamefont {A.~Y.}\ \bibnamefont
  {Kitaev}},\ }\href {\doibase 10.1006/aphy.2002.6254} {\bibfield  {journal}
  {\bibinfo  {journal} {Ann. Phys. (N. Y.)}\ }\textbf {\bibinfo {volume}
  {298}},\ \bibinfo {pages} {210 } (\bibinfo {year} {2002})}\BibitemShut
  {NoStop}%
\bibitem [{\citenamefont {Bravyi}(2006)}]{bravyi2006universal}%
  \BibitemOpen
  \bibfield  {author} {\bibinfo {author} {\bibfnamefont {S.}~\bibnamefont
  {Bravyi}},\ }\href {\doibase 10.1103/PhysRevA.73.042313} {\bibfield
  {journal} {\bibinfo  {journal} {Phys. Rev. A}\ }\textbf {\bibinfo {volume}
  {73}},\ \bibinfo {pages} {042313} (\bibinfo {year} {2006})}\BibitemShut
  {NoStop}%
\bibitem [{\citenamefont {Jozsa}\ and\ \citenamefont
  {Miyake}(2008)}]{jozsa2008matchgates}%
  \BibitemOpen
  \bibfield  {author} {\bibinfo {author} {\bibfnamefont {R.}~\bibnamefont
  {Jozsa}}\ and\ \bibinfo {author} {\bibfnamefont {A.}~\bibnamefont {Miyake}},\
  }\href {\doibase 10.1098/rspa.2008.0189} {\bibfield  {journal} {\bibinfo
  {journal} {Proceedings of the Royal Society A: Mathematical, Physical and
  Engineering Sciences}\ }\textbf {\bibinfo {volume} {464}},\ \bibinfo {pages}
  {3089} (\bibinfo {year} {2008})}\BibitemShut {NoStop}%
\bibitem [{\citenamefont {de~Melo}\ \emph {et~al.}(2013)\citenamefont
  {de~Melo}, \citenamefont {{{\'C}wikli{\'n}ski}},\ and\ \citenamefont
  {Terhal}}]{melo2013power}%
  \BibitemOpen
  \bibfield  {author} {\bibinfo {author} {\bibfnamefont {F.}~\bibnamefont
  {de~Melo}}, \bibinfo {author} {\bibfnamefont {P.}~\bibnamefont
  {{{\'C}wikli{\'n}ski}}}, \ and\ \bibinfo {author} {\bibfnamefont {B.~M.}\
  \bibnamefont {Terhal}},\ }\href
  {http://stacks.iop.org/1367-2630/15/i=1/a=013015} {\bibfield  {journal}
  {\bibinfo  {journal} {New J. Phys.}\ }\textbf {\bibinfo {volume} {15}},\
  \bibinfo {pages} {013015} (\bibinfo {year} {2013})}\BibitemShut {NoStop}%
\bibitem [{\citenamefont {{Brod}}\ and\ \citenamefont
  {{Childs}}(2013)}]{brod2014computational}%
  \BibitemOpen
  \bibfield  {author} {\bibinfo {author} {\bibfnamefont {D.~J.}\ \bibnamefont
  {{Brod}}}\ and\ \bibinfo {author} {\bibfnamefont {A.~M.}\ \bibnamefont
  {{Childs}}},\ }\href@noop {} {\bibfield  {journal} {\bibinfo  {journal}
  {ArXiv e-prints}\ } (\bibinfo {year} {2013})},\ \Eprint
  {http://arxiv.org/abs/1308.1463} {arXiv:1308.1463 [quant-ph]} \BibitemShut
  {NoStop}%
\bibitem [{\citenamefont {{Kitaev}}(2001)}]{kitaev2001unpaired}%
  \BibitemOpen
  \bibfield  {author} {\bibinfo {author} {\bibfnamefont {A.~Y.}\ \bibnamefont
  {{Kitaev}}},\ }\href {\doibase 10.1070/1063-7869/44/10S/S29} {\bibfield
  {journal} {\bibinfo  {journal} {Physics Uspekhi}\ }\textbf {\bibinfo {volume}
  {44}},\ \bibinfo {pages} {131} (\bibinfo {year} {2001})},\ \Eprint
  {http://arxiv.org/abs/cond-mat/0010440} {arXiv:cond-mat/0010440
  [cond-mat.mes-hall]} \BibitemShut {NoStop}%
\bibitem [{\citenamefont {Bravyi}\ \emph {et~al.}(2010)\citenamefont {Bravyi},
  \citenamefont {Terhal},\ and\ \citenamefont {Leemhuis}}]{bravyi2010majorana}%
  \BibitemOpen
  \bibfield  {author} {\bibinfo {author} {\bibfnamefont {S.}~\bibnamefont
  {Bravyi}}, \bibinfo {author} {\bibfnamefont {B.~M.}\ \bibnamefont {Terhal}},
  \ and\ \bibinfo {author} {\bibfnamefont {B.}~\bibnamefont {Leemhuis}},\
  }\href {\doibase 10.1088/1367-2630/12/8/083039} {\ \textbf {\bibinfo {volume}
  {12}},\ \bibinfo {pages} {083039} (\bibinfo {year} {2010})}\BibitemShut
  {NoStop}%
\bibitem [{\citenamefont {{Hastings}}(2017)}]{hastings2017small}%
  \BibitemOpen
  \bibfield  {author} {\bibinfo {author} {\bibfnamefont {M.~B.}\ \bibnamefont
  {{Hastings}}},\ }\href@noop {} {\bibfield  {journal} {\bibinfo  {journal}
  {arXiv e-prints}\ ,\ \bibinfo {eid} {arXiv:1703.00612}} (\bibinfo {year}
  {2017})},\ \Eprint {http://arxiv.org/abs/1703.00612} {arXiv:1703.00612
  [quant-ph]} \BibitemShut {NoStop}%
\bibitem [{\citenamefont {{Vijay}}\ and\ \citenamefont
  {{Fu}}(2017)}]{vijay2017quantum}%
  \BibitemOpen
  \bibfield  {author} {\bibinfo {author} {\bibfnamefont {S.}~\bibnamefont
  {{Vijay}}}\ and\ \bibinfo {author} {\bibfnamefont {L.}~\bibnamefont {{Fu}}},\
  }\href@noop {} {\bibfield  {journal} {\bibinfo  {journal} {arXiv e-prints}\
  ,\ \bibinfo {eid} {arXiv:1703.00459}} (\bibinfo {year} {2017})},\ \Eprint
  {http://arxiv.org/abs/1703.00459} {arXiv:1703.00459 [cond-mat.mes-hall]}
  \BibitemShut {NoStop}%
\bibitem [{\citenamefont {{Viyuela}}\ \emph {et~al.}(2019)\citenamefont
  {{Viyuela}}, \citenamefont {{Vijay}},\ and\ \citenamefont
  {{Fu}}}]{viyuela2019scalable}%
  \BibitemOpen
  \bibfield  {author} {\bibinfo {author} {\bibfnamefont {O.}~\bibnamefont
  {{Viyuela}}}, \bibinfo {author} {\bibfnamefont {S.}~\bibnamefont {{Vijay}}},
  \ and\ \bibinfo {author} {\bibfnamefont {L.}~\bibnamefont {{Fu}}},\ }\href
  {\doibase 10.1103/PhysRevB.99.205114} {\bibfield  {journal} {\bibinfo
  {journal} {\prb}\ }\textbf {\bibinfo {volume} {99}},\ \bibinfo {eid} {205114}
  (\bibinfo {year} {2019})},\ \Eprint {http://arxiv.org/abs/1812.08477}
  {arXiv:1812.08477 [quant-ph]} \BibitemShut {NoStop}%
\bibitem [{\citenamefont {Chapman}\ and\ \citenamefont
  {Flammia}(2020)}]{chapman2020characterization}%
  \BibitemOpen
  \bibfield  {author} {\bibinfo {author} {\bibfnamefont {A.}~\bibnamefont
  {Chapman}}\ and\ \bibinfo {author} {\bibfnamefont {S.~T.}\ \bibnamefont
  {Flammia}},\ }\href {\doibase 10.22331/q-2020-06-04-278} {\bibfield
  {journal} {\bibinfo  {journal} {Quantum}\ }\textbf {\bibinfo {volume} {4}},\
  \bibinfo {pages} {278} (\bibinfo {year} {2020})}\BibitemShut {NoStop}%
\bibitem [{\citenamefont {{Ogura}}\ \emph {et~al.}(2020)\citenamefont
  {{Ogura}}, \citenamefont {{Imamura}}, \citenamefont {{Kameyama}},
  \citenamefont {{Minami}},\ and\ \citenamefont {{Sato}}}]{ogura2020geometric}%
  \BibitemOpen
  \bibfield  {author} {\bibinfo {author} {\bibfnamefont {M.}~\bibnamefont
  {{Ogura}}}, \bibinfo {author} {\bibfnamefont {Y.}~\bibnamefont {{Imamura}}},
  \bibinfo {author} {\bibfnamefont {N.}~\bibnamefont {{Kameyama}}}, \bibinfo
  {author} {\bibfnamefont {K.}~\bibnamefont {{Minami}}}, \ and\ \bibinfo
  {author} {\bibfnamefont {M.}~\bibnamefont {{Sato}}},\ }\href {\doibase
  10.1103/PhysRevB.102.245118} {\bibfield  {journal} {\bibinfo  {journal}
  {\prb}\ }\textbf {\bibinfo {volume} {102}},\ \bibinfo {eid} {245118}
  (\bibinfo {year} {2020})},\ \Eprint {http://arxiv.org/abs/2003.13264}
  {arXiv:2003.13264 [cond-mat.stat-mech]} \BibitemShut {NoStop}%
\bibitem [{\citenamefont {Verstraete}\ and\ \citenamefont
  {Cirac}(2005)}]{verstraete2005mapping}%
  \BibitemOpen
  \bibfield  {author} {\bibinfo {author} {\bibfnamefont {F.}~\bibnamefont
  {Verstraete}}\ and\ \bibinfo {author} {\bibfnamefont {J.~I.}\ \bibnamefont
  {Cirac}},\ }\href {\doibase 10.1088/1742-5468/2005/09/p09012} {\bibfield
  {journal} {\bibinfo  {journal} {Journal of Statistical Mechanics: Theory and
  Experiment}\ }\textbf {\bibinfo {volume} {2005}},\ \bibinfo {pages} {P09012}
  (\bibinfo {year} {2005})}\BibitemShut {NoStop}%
\bibitem [{\citenamefont {{Setia}}\ \emph {et~al.}(2018)\citenamefont
  {{Setia}}, \citenamefont {{Bravyi}}, \citenamefont {{Mezzacapo}},\ and\
  \citenamefont {{Whitfield}}}]{setia2018superfast}%
  \BibitemOpen
  \bibfield  {author} {\bibinfo {author} {\bibfnamefont {K.}~\bibnamefont
  {{Setia}}}, \bibinfo {author} {\bibfnamefont {S.}~\bibnamefont {{Bravyi}}},
  \bibinfo {author} {\bibfnamefont {A.}~\bibnamefont {{Mezzacapo}}}, \ and\
  \bibinfo {author} {\bibfnamefont {J.~D.}\ \bibnamefont {{Whitfield}}},\
  }\href@noop {} {\bibfield  {journal} {\bibinfo  {journal} {arXiv e-prints}\
  ,\ \bibinfo {eid} {arXiv:1810.05274}} (\bibinfo {year} {2018})},\ \Eprint
  {http://arxiv.org/abs/1810.05274} {arXiv:1810.05274 [quant-ph]} \BibitemShut
  {NoStop}%
\bibitem [{\citenamefont {Seeley}\ \emph {et~al.}(2012)\citenamefont {Seeley},
  \citenamefont {Richard},\ and\ \citenamefont
  {Love}}]{seeley2012bravyikitaev}%
  \BibitemOpen
  \bibfield  {author} {\bibinfo {author} {\bibfnamefont {J.~T.}\ \bibnamefont
  {Seeley}}, \bibinfo {author} {\bibfnamefont {M.~J.}\ \bibnamefont {Richard}},
  \ and\ \bibinfo {author} {\bibfnamefont {P.~J.}\ \bibnamefont {Love}},\
  }\href {\doibase 10.1063/1.4768229} {\bibfield  {journal} {\bibinfo
  {journal} {The Journal of Chemical Physics}\ }\textbf {\bibinfo {volume}
  {137}},\ \bibinfo {pages} {224109} (\bibinfo {year} {2012})},\ \Eprint
  {http://arxiv.org/abs/https://doi.org/10.1063/1.4768229}
  {https://doi.org/10.1063/1.4768229} \BibitemShut {NoStop}%
\bibitem [{\citenamefont {{Bravyi}}\ \emph {et~al.}(2017)\citenamefont
  {{Bravyi}}, \citenamefont {{Gambetta}}, \citenamefont {{Mezzacapo}},\ and\
  \citenamefont {{Temme}}}]{bravyi2017tapering}%
  \BibitemOpen
  \bibfield  {author} {\bibinfo {author} {\bibfnamefont {S.}~\bibnamefont
  {{Bravyi}}}, \bibinfo {author} {\bibfnamefont {J.~M.}\ \bibnamefont
  {{Gambetta}}}, \bibinfo {author} {\bibfnamefont {A.}~\bibnamefont
  {{Mezzacapo}}}, \ and\ \bibinfo {author} {\bibfnamefont {K.}~\bibnamefont
  {{Temme}}},\ }\href@noop {} {\bibfield  {journal} {\bibinfo  {journal} {arXiv
  e-prints}\ ,\ \bibinfo {eid} {arXiv:1701.08213}} (\bibinfo {year} {2017})},\
  \Eprint {http://arxiv.org/abs/1701.08213} {arXiv:1701.08213 [quant-ph]}
  \BibitemShut {NoStop}%
\bibitem [{\citenamefont {Steudtner}\ and\ \citenamefont
  {Wehner}(2018)}]{steudtner2018fermion}%
  \BibitemOpen
  \bibfield  {author} {\bibinfo {author} {\bibfnamefont {M.}~\bibnamefont
  {Steudtner}}\ and\ \bibinfo {author} {\bibfnamefont {S.}~\bibnamefont
  {Wehner}},\ }\href {\doibase 10.1088/1367-2630/aac54f} {\bibfield  {journal}
  {\bibinfo  {journal} {New Journal of Physics}\ }\textbf {\bibinfo {volume}
  {20}},\ \bibinfo {pages} {063010} (\bibinfo {year} {2018})}\BibitemShut
  {NoStop}%
\bibitem [{\citenamefont {{Jiang}}\ \emph
  {et~al.}(2019{\natexlab{a}})\citenamefont {{Jiang}}, \citenamefont {{Kalev}},
  \citenamefont {{Mruczkiewicz}},\ and\ \citenamefont
  {{Neven}}}]{jiang2019optimal}%
  \BibitemOpen
  \bibfield  {author} {\bibinfo {author} {\bibfnamefont {Z.}~\bibnamefont
  {{Jiang}}}, \bibinfo {author} {\bibfnamefont {A.}~\bibnamefont {{Kalev}}},
  \bibinfo {author} {\bibfnamefont {W.}~\bibnamefont {{Mruczkiewicz}}}, \ and\
  \bibinfo {author} {\bibfnamefont {H.}~\bibnamefont {{Neven}}},\ }\href@noop
  {} {\bibfield  {journal} {\bibinfo  {journal} {arXiv e-prints}\ ,\ \bibinfo
  {eid} {arXiv:1910.10746}} (\bibinfo {year} {2019}{\natexlab{a}})},\ \Eprint
  {http://arxiv.org/abs/1910.10746} {arXiv:1910.10746 [quant-ph]} \BibitemShut
  {NoStop}%
\bibitem [{\citenamefont {Derby}\ \emph {et~al.}(2021)\citenamefont {Derby},
  \citenamefont {Klassen}, \citenamefont {Bausch},\ and\ \citenamefont
  {Cubitt}}]{derby2021compact}%
  \BibitemOpen
  \bibfield  {author} {\bibinfo {author} {\bibfnamefont {C.}~\bibnamefont
  {Derby}}, \bibinfo {author} {\bibfnamefont {J.}~\bibnamefont {Klassen}},
  \bibinfo {author} {\bibfnamefont {J.}~\bibnamefont {Bausch}}, \ and\ \bibinfo
  {author} {\bibfnamefont {T.}~\bibnamefont {Cubitt}},\ }\href {\doibase
  10.1103/PhysRevB.104.035118} {\bibfield  {journal} {\bibinfo  {journal}
  {Phys. Rev. B}\ }\textbf {\bibinfo {volume} {104}},\ \bibinfo {pages}
  {035118} (\bibinfo {year} {2021})}\BibitemShut {NoStop}%
\bibitem [{\citenamefont {{Chiew}}\ and\ \citenamefont
  {{Strelchuk}}(2021)}]{chiew2021optimal}%
  \BibitemOpen
  \bibfield  {author} {\bibinfo {author} {\bibfnamefont {M.}~\bibnamefont
  {{Chiew}}}\ and\ \bibinfo {author} {\bibfnamefont {S.}~\bibnamefont
  {{Strelchuk}}},\ }\href@noop {} {\bibfield  {journal} {\bibinfo  {journal}
  {arXiv e-prints}\ ,\ \bibinfo {eid} {arXiv:2110.12792}} (\bibinfo {year}
  {2021})},\ \Eprint {http://arxiv.org/abs/2110.12792} {arXiv:2110.12792
  [quant-ph]} \BibitemShut {NoStop}%
\bibitem [{\citenamefont {{Jiang}}\ \emph
  {et~al.}(2019{\natexlab{b}})\citenamefont {{Jiang}}, \citenamefont
  {{McClean}}, \citenamefont {{Babbush}},\ and\ \citenamefont
  {{Neven}}}]{jiang2019majorana}%
  \BibitemOpen
  \bibfield  {author} {\bibinfo {author} {\bibfnamefont {Z.}~\bibnamefont
  {{Jiang}}}, \bibinfo {author} {\bibfnamefont {J.}~\bibnamefont {{McClean}}},
  \bibinfo {author} {\bibfnamefont {R.}~\bibnamefont {{Babbush}}}, \ and\
  \bibinfo {author} {\bibfnamefont {H.}~\bibnamefont {{Neven}}},\ }\href
  {\doibase 10.1103/PhysRevApplied.12.064041} {\bibfield  {journal} {\bibinfo
  {journal} {Physical Review Applied}\ }\textbf {\bibinfo {volume} {12}},\
  \bibinfo {eid} {064041} (\bibinfo {year} {2019}{\natexlab{b}})},\ \Eprint
  {http://arxiv.org/abs/1812.08190} {arXiv:1812.08190 [quant-ph]} \BibitemShut
  {NoStop}%
\bibitem [{\citenamefont {Haah}(2013)}]{haah2013commuting}%
  \BibitemOpen
  \bibfield  {author} {\bibinfo {author} {\bibfnamefont {J.}~\bibnamefont
  {Haah}},\ }\href {\doibase 10.1007/s00220-013-1810-2} {\bibfield  {journal}
  {\bibinfo  {journal} {Communications in Mathematical Physics}\ }\textbf
  {\bibinfo {volume} {324}},\ \bibinfo {pages} {351} (\bibinfo {year}
  {2013})}\BibitemShut {NoStop}%
\bibitem [{\citenamefont {{Yu}}\ \emph {et~al.}(2008)\citenamefont {{Yu}},
  \citenamefont {{Kou}},\ and\ \citenamefont {{Wen}}}]{Yu2008Topological}%
  \BibitemOpen
  \bibfield  {author} {\bibinfo {author} {\bibfnamefont {J.}~\bibnamefont
  {{Yu}}}, \bibinfo {author} {\bibfnamefont {S.-P.}\ \bibnamefont {{Kou}}}, \
  and\ \bibinfo {author} {\bibfnamefont {X.-G.}\ \bibnamefont {{Wen}}},\ }\href
  {\doibase 10.1209/0295-5075/84/17004} {\bibfield  {journal} {\bibinfo
  {journal} {EPL (Europhysics Letters)}\ }\textbf {\bibinfo {volume} {84}},\
  \bibinfo {pages} {17004} (\bibinfo {year} {2008})}\BibitemShut {NoStop}%
\bibitem [{\citenamefont {{Bravyi}}\ \emph {et~al.}(2012)\citenamefont
  {{Bravyi}}, \citenamefont {{Duclos-Cianci}}, \citenamefont {{Poulin}},\ and\
  \citenamefont {{Suchara}}}]{bravyi2012subsystem}%
  \BibitemOpen
  \bibfield  {author} {\bibinfo {author} {\bibfnamefont {S.}~\bibnamefont
  {{Bravyi}}}, \bibinfo {author} {\bibfnamefont {G.}~\bibnamefont
  {{Duclos-Cianci}}}, \bibinfo {author} {\bibfnamefont {D.}~\bibnamefont
  {{Poulin}}}, \ and\ \bibinfo {author} {\bibfnamefont {M.}~\bibnamefont
  {{Suchara}}},\ }\href@noop {} {\bibfield  {journal} {\bibinfo  {journal}
  {arXiv e-prints}\ ,\ \bibinfo {eid} {arXiv:1207.1443}} (\bibinfo {year}
  {2012})},\ \Eprint {http://arxiv.org/abs/1207.1443} {arXiv:1207.1443
  [quant-ph]} \BibitemShut {NoStop}%
\bibitem [{\citenamefont {Bacon}(2006)}]{Bacon2006}%
  \BibitemOpen
  \bibfield  {author} {\bibinfo {author} {\bibfnamefont {D.}~\bibnamefont
  {Bacon}},\ }\href {\doibase 10.1103/physreva.73.012340} {\bibfield  {journal}
  {\bibinfo  {journal} {Physical Review A}\ }\textbf {\bibinfo {volume} {73}},\
  \bibinfo {pages} {012340} (\bibinfo {year} {2006})},\ \Eprint
  {http://arxiv.org/abs/quant-ph/0506023} {quant-ph/0506023} \BibitemShut
  {NoStop}%
\bibitem [{\citenamefont {{Bombin}}(2010)}]{Bombin2010Topological}%
  \BibitemOpen
  \bibfield  {author} {\bibinfo {author} {\bibfnamefont {H.}~\bibnamefont
  {{Bombin}}},\ }\href {\doibase 10.1103/PhysRevA.81.032301} {\bibfield
  {journal} {\bibinfo  {journal} {\pra}\ }\textbf {\bibinfo {volume} {81}},\
  \bibinfo {eid} {032301} (\bibinfo {year} {2010})},\ \Eprint
  {http://arxiv.org/abs/0908.4246} {arXiv:0908.4246 [quant-ph]} \BibitemShut
  {NoStop}%
\bibitem [{\citenamefont {{Suchara}}\ \emph {et~al.}(2011)\citenamefont
  {{Suchara}}, \citenamefont {{Bravyi}},\ and\ \citenamefont
  {{Terhal}}}]{suchara2011constructions}%
  \BibitemOpen
  \bibfield  {author} {\bibinfo {author} {\bibfnamefont {M.}~\bibnamefont
  {{Suchara}}}, \bibinfo {author} {\bibfnamefont {S.}~\bibnamefont {{Bravyi}}},
  \ and\ \bibinfo {author} {\bibfnamefont {B.}~\bibnamefont {{Terhal}}},\
  }\href {\doibase 10.1088/1751-8113/44/15/155301} {\bibfield  {journal}
  {\bibinfo  {journal} {Journal of Physics A Mathematical General}\ }\textbf
  {\bibinfo {volume} {44}},\ \bibinfo {eid} {155301} (\bibinfo {year}
  {2011})},\ \Eprint {http://arxiv.org/abs/1012.0425} {arXiv:1012.0425
  [quant-ph]} \BibitemShut {NoStop}%
\bibitem [{\citenamefont {Adiga}\ \emph {et~al.}(2010)\citenamefont {Adiga},
  \citenamefont {Balakrishnan},\ and\ \citenamefont {So}}]{adiga2010skew}%
  \BibitemOpen
  \bibfield  {author} {\bibinfo {author} {\bibfnamefont {C.}~\bibnamefont
  {Adiga}}, \bibinfo {author} {\bibfnamefont {R.}~\bibnamefont {Balakrishnan}},
  \ and\ \bibinfo {author} {\bibfnamefont {W.}~\bibnamefont {So}},\ }\href
  {\doibase 10.1016/j.laa.2009.11.034} {\bibfield  {journal} {\bibinfo
  {journal} {Linear Algebra and its Applications}\ }\textbf {\bibinfo {volume}
  {432}},\ \bibinfo {pages} {1825} (\bibinfo {year} {2010})}\BibitemShut
  {NoStop}%
\bibitem [{\citenamefont {{Poulin}}(2005)}]{poulin2005stabilizer}%
  \BibitemOpen
  \bibfield  {author} {\bibinfo {author} {\bibfnamefont {D.}~\bibnamefont
  {{Poulin}}},\ }\href {\doibase 10.1103/PhysRevLett.95.230504} {\bibfield
  {journal} {\bibinfo  {journal} {\prl}\ }\textbf {\bibinfo {volume} {95}},\
  \bibinfo {eid} {230504} (\bibinfo {year} {2005})},\ \Eprint
  {http://arxiv.org/abs/quant-ph/0508131} {arXiv:quant-ph/0508131 [quant-ph]}
  \BibitemShut {NoStop}%
\bibitem [{\citenamefont {Nielsen}\ and\ \citenamefont
  {Chuang}(2011)}]{nielsen2011quantum}%
  \BibitemOpen
  \bibfield  {author} {\bibinfo {author} {\bibfnamefont {M.~A.}\ \bibnamefont
  {Nielsen}}\ and\ \bibinfo {author} {\bibfnamefont {I.~L.}\ \bibnamefont
  {Chuang}},\ }\href@noop {} {\emph {\bibinfo {title} {Quantum Computation and
  Quantum Information: 10th Anniversary Edition}}},\ \bibinfo {edition} {10th}\
  ed.\ (\bibinfo  {publisher} {Cambridge University Press},\ \bibinfo {address}
  {USA},\ \bibinfo {year} {2011})\BibitemShut {NoStop}%
\bibitem [{\citenamefont
  {{Tantivasadakarn}}(2020)}]{tantivasadakarn2020jordanwigner}%
  \BibitemOpen
  \bibfield  {author} {\bibinfo {author} {\bibfnamefont {N.}~\bibnamefont
  {{Tantivasadakarn}}},\ }\href@noop {} {\bibfield  {journal} {\bibinfo
  {journal} {arXiv e-prints}\ ,\ \bibinfo {eid} {arXiv:2002.11345}} (\bibinfo
  {year} {2020})},\ \Eprint {http://arxiv.org/abs/2002.11345} {arXiv:2002.11345
  [cond-mat.str-el]} \BibitemShut {NoStop}%
\bibitem [{\citenamefont {Chen}\ \emph {et~al.}(2018)\citenamefont {Chen},
  \citenamefont {Kapustin},\ and\ \citenamefont {Radi{\v c}evi{\'
  c}}}]{chen2018exact}%
  \BibitemOpen
  \bibfield  {author} {\bibinfo {author} {\bibfnamefont {Y.-A.}\ \bibnamefont
  {Chen}}, \bibinfo {author} {\bibfnamefont {A.}~\bibnamefont {Kapustin}}, \
  and\ \bibinfo {author} {\bibfnamefont {{\DH}.}~\bibnamefont {Radi{\v c}evi{\'
  c}}},\ }\href {\doibase https://doi.org/10.1016/j.aop.2018.03.024} {\bibfield
   {journal} {\bibinfo  {journal} {Annals of Physics}\ }\textbf {\bibinfo
  {volume} {393}},\ \bibinfo {pages} {234 } (\bibinfo {year}
  {2018})}\BibitemShut {NoStop}%
\bibitem [{\citenamefont {Shor}(1995)}]{Shor1995}%
  \BibitemOpen
  \bibfield  {author} {\bibinfo {author} {\bibfnamefont {P.~W.}\ \bibnamefont
  {Shor}},\ }\href {\doibase 10.1103/physreva.52.r2493} {\bibfield  {journal}
  {\bibinfo  {journal} {Physical Review A}\ }\textbf {\bibinfo {volume} {52}},\
  \bibinfo {pages} {R2493} (\bibinfo {year} {1995})}\BibitemShut {NoStop}%
\bibitem [{\citenamefont {Napp}\ and\ \citenamefont
  {Prreskill}(2013)}]{NappPreskill2013}%
  \BibitemOpen
  \bibfield  {author} {\bibinfo {author} {\bibfnamefont {J.}~\bibnamefont
  {Napp}}\ and\ \bibinfo {author} {\bibfnamefont {J.}~\bibnamefont
  {Prreskill}},\ }\href@noop {} {\bibfield  {journal} {\bibinfo  {journal}
  {Quantum Information and Computation}\ }\textbf {\bibinfo {volume} {13}},\
  \bibinfo {pages} {0490} (\bibinfo {year} {2013})},\ \Eprint
  {http://arxiv.org/abs/1209.0794} {1209.0794} \BibitemShut {NoStop}%
\bibitem [{\citenamefont {Kitaev}(2003)}]{Kitaev:2003toric}%
  \BibitemOpen
  \bibfield  {author} {\bibinfo {author} {\bibfnamefont {A.~Y.}\ \bibnamefont
  {Kitaev}},\ }\href@noop {} {\bibfield  {journal} {\bibinfo  {journal} {Annals
  of Physics}\ }\textbf {\bibinfo {volume} {303}},\ \bibinfo {pages} {1 }
  (\bibinfo {year} {2003})}\BibitemShut {NoStop}%
\bibitem [{\citenamefont {Elman}\ \emph {et~al.}(2021)\citenamefont {Elman},
  \citenamefont {Chapman},\ and\ \citenamefont {Flammia}}]{elman2021free}%
  \BibitemOpen
  \bibfield  {author} {\bibinfo {author} {\bibfnamefont {S.~J.}\ \bibnamefont
  {Elman}}, \bibinfo {author} {\bibfnamefont {A.}~\bibnamefont {Chapman}}, \
  and\ \bibinfo {author} {\bibfnamefont {S.~T.}\ \bibnamefont {Flammia}},\
  }\href {\doibase 10.1007/s00220-021-04220-w} {\bibfield  {journal} {\bibinfo
  {journal} {Communications in Mathematical Physics}\ }\textbf {\bibinfo
  {volume} {388}},\ \bibinfo {pages} {969} (\bibinfo {year}
  {2021})}\BibitemShut {NoStop}%
\bibitem [{\citenamefont {Lieb}(1994)}]{lieb1994flux}%
  \BibitemOpen
  \bibfield  {author} {\bibinfo {author} {\bibfnamefont {E.~H.}\ \bibnamefont
  {Lieb}},\ }\href {\doibase 10.1103/PhysRevLett.73.2158} {\bibfield  {journal}
  {\bibinfo  {journal} {Phys. Rev. Lett.}\ }\textbf {\bibinfo {volume} {73}},\
  \bibinfo {pages} {2158} (\bibinfo {year} {1994})}\BibitemShut {NoStop}%
\bibitem [{\citenamefont {Whitney}(1932)}]{whitney1932congruent}%
  \BibitemOpen
  \bibfield  {author} {\bibinfo {author} {\bibfnamefont {H.}~\bibnamefont
  {Whitney}},\ }\href {http://www.jstor.org/stable/2371086} {\bibfield
  {journal} {\bibinfo  {journal} {American Journal of Mathematics}\ }\textbf
  {\bibinfo {volume} {54}},\ \bibinfo {pages} {150} (\bibinfo {year}
  {1932})}\BibitemShut {NoStop}%
\bibitem [{\citenamefont {Krausz}(1943)}]{krausz1943demonstration}%
  \BibitemOpen
  \bibfield  {author} {\bibinfo {author} {\bibfnamefont {J.}~\bibnamefont
  {Krausz}},\ }\href {http://real-j.mtak.hu/7300/} {\bibfield  {journal}
  {\bibinfo  {journal} {Matematikai {\'e}s Fizikai Lapok}\ }\textbf {\bibinfo
  {volume} {50}} (\bibinfo {year} {1943})}\BibitemShut {NoStop}%
\bibitem [{\citenamefont {{Li}}\ and\ \citenamefont
  {{Lian}}(2013)}]{li2013survey}%
  \BibitemOpen
  \bibfield  {author} {\bibinfo {author} {\bibfnamefont {X.}~\bibnamefont
  {{Li}}}\ and\ \bibinfo {author} {\bibfnamefont {H.}~\bibnamefont {{Lian}}},\
  }\href@noop {} {\bibfield  {journal} {\bibinfo  {journal} {arXiv e-prints}\
  ,\ \bibinfo {eid} {arXiv:1304.5707}} (\bibinfo {year} {2013})},\ \Eprint
  {http://arxiv.org/abs/1304.5707} {arXiv:1304.5707 [math.CO]} \BibitemShut
  {NoStop}%
\bibitem [{\citenamefont {Denglan}\ and\ \citenamefont
  {Yaoping}(2013)}]{denglan2013skew}%
  \BibitemOpen
  \bibfield  {author} {\bibinfo {author} {\bibfnamefont {C.}~\bibnamefont
  {Denglan}}\ and\ \bibinfo {author} {\bibfnamefont {H.}~\bibnamefont
  {Yaoping}},\ }\href {\doibase 10.37236/2864} {\bibfield  {journal} {\bibinfo
  {journal} {The Electronic Journal of Combinatorics}\ }\textbf {\bibinfo
  {volume} {20}} (\bibinfo {year} {2013}),\ 10.37236/2864}\BibitemShut
  {NoStop}%
\bibitem [{\citenamefont {Koll\'{a}r}\ and\ \citenamefont
  {Sarnak}(2021)}]{kollar2020gap}%
  \BibitemOpen
  \bibfield  {author} {\bibinfo {author} {\bibfnamefont {A.~J.}\ \bibnamefont
  {Koll\'{a}r}}\ and\ \bibinfo {author} {\bibfnamefont {P.}~\bibnamefont
  {Sarnak}},\ }\href {\doibase 10.1090/cams/3} {\bibfield  {journal} {\bibinfo
  {journal} {Communications of the American Mathematical Society}\ }\textbf
  {\bibinfo {volume} {1}},\ \bibinfo {pages} {1} (\bibinfo {year}
  {2021})}\BibitemShut {NoStop}%
\bibitem [{\citenamefont {Biggs}(1993)}]{BiggsGraphTheory}%
  \BibitemOpen
  \bibfield  {author} {\bibinfo {author} {\bibfnamefont {N.}~\bibnamefont
  {Biggs}},\ }\href@noop {} {\emph {\bibinfo {title} {{Algebraic Graph
  Theory}}}},\ \bibinfo {edition} {2nd}\ ed.\ (\bibinfo  {publisher} {Cambridge
  University Press},\ \bibinfo {address} {Cambridge},\ \bibinfo {year}
  {1993})\BibitemShut {NoStop}%
\bibitem [{\citenamefont {Cvetkovi{\'c}}\ \emph {et~al.}(1980)\citenamefont
  {Cvetkovi{\'c}}, \citenamefont {Doob},\ and\ \citenamefont {Sachs}}]{CDS}%
  \BibitemOpen
  \bibfield  {author} {\bibinfo {author} {\bibfnamefont {D.~M.}\ \bibnamefont
  {Cvetkovi{\'c}}}, \bibinfo {author} {\bibfnamefont {M.}~\bibnamefont {Doob}},
  \ and\ \bibinfo {author} {\bibfnamefont {H.}~\bibnamefont {Sachs}},\
  }\href@noop {} {\emph {\bibinfo {title} {{Spectra of Graphs: Theory and
  Application}}}}\ (\bibinfo  {publisher} {Academic Press},\ \bibinfo {year}
  {1980})\BibitemShut {NoStop}%
\bibitem [{\citenamefont {Yaoping}\ \emph {et~al.}(2011)\citenamefont
  {Yaoping}, \citenamefont {Xiaoling},\ and\ \citenamefont
  {Chongyan}}]{yaoping2011oriented}%
  \BibitemOpen
  \bibfield  {author} {\bibinfo {author} {\bibfnamefont {H.}~\bibnamefont
  {Yaoping}}, \bibinfo {author} {\bibfnamefont {S.}~\bibnamefont {Xiaoling}}, \
  and\ \bibinfo {author} {\bibfnamefont {Z.}~\bibnamefont {Chongyan}},\
  }\href@noop {} {\enquote {\bibinfo {title} {Oriented unicyclic graphs with
  extremal skew energy},}\ } (\bibinfo {year} {2011}),\ \Eprint
  {http://arxiv.org/abs/1108.6229} {arXiv:1108.6229 [math.CO]} \BibitemShut
  {NoStop}%
\bibitem [{\citenamefont {Guo}\ and\ \citenamefont {Mohar}(2014)}]{MoharGap}%
  \BibitemOpen
  \bibfield  {author} {\bibinfo {author} {\bibfnamefont {K.}~\bibnamefont
  {Guo}}\ and\ \bibinfo {author} {\bibfnamefont {B.}~\bibnamefont {Mohar}},\
  }\href {\doibase 10.1016/j.laa.2014.02.016} {\bibfield  {journal} {\bibinfo
  {journal} {Linear Alg. Appl.}\ }\textbf {\bibinfo {volume} {449}},\ \bibinfo
  {pages} {68} (\bibinfo {year} {2014})}\BibitemShut {NoStop}%
\bibitem [{\citenamefont {Mohar}(2016)}]{MoharMedian}%
  \BibitemOpen
  \bibfield  {author} {\bibinfo {author} {\bibfnamefont {B.}~\bibnamefont
  {Mohar}},\ }\href {\doibase 10.1017/s0963548316000201} {\bibfield  {journal}
  {\bibinfo  {journal} {Comb. Prob. Comput.}\ }\textbf {\bibinfo {volume}
  {25}},\ \bibinfo {pages} {768} (\bibinfo {year} {2016})},\ \Eprint
  {http://arxiv.org/abs/1309.7395} {1309.7395} \BibitemShut {NoStop}%
\bibitem [{\citenamefont {Girvin}\ and\ \citenamefont {Yang}(2019)}]{GirvinSS}%
  \BibitemOpen
  \bibfield  {author} {\bibinfo {author} {\bibfnamefont {S.~M.}\ \bibnamefont
  {Girvin}}\ and\ \bibinfo {author} {\bibfnamefont {K.}~\bibnamefont {Yang}},\
  }\href@noop {} {\emph {\bibinfo {title} {{Modern Condensed Matter
  Physics}}}}\ (\bibinfo  {publisher} {Cambridge University Press, Inc.},\
  \bibinfo {year} {2019})\BibitemShut {NoStop}%
\bibitem [{\citenamefont {Koll\'{a}r}\ \emph {et~al.}(2019)\citenamefont
  {Koll\'{a}r}, \citenamefont {Fitzpatrick}, \citenamefont {Sarnak},\ and\
  \citenamefont {Houck}}]{Kollar:2019linegraph}%
  \BibitemOpen
  \bibfield  {author} {\bibinfo {author} {\bibfnamefont {A.~J.}\ \bibnamefont
  {Koll\'{a}r}}, \bibinfo {author} {\bibfnamefont {M.}~\bibnamefont
  {Fitzpatrick}}, \bibinfo {author} {\bibfnamefont {P.}~\bibnamefont {Sarnak}},
  \ and\ \bibinfo {author} {\bibfnamefont {A.~A.}\ \bibnamefont {Houck}},\
  }\href {\doibase 10.1016/0012-365x(87)90167-1} {\bibfield  {journal}
  {\bibinfo  {journal} {Commun. Math. Phys.}\ }\textbf {\bibinfo {volume}
  {44}},\ \bibinfo {pages} {1601} (\bibinfo {year} {2019})}\BibitemShut
  {NoStop}%
\bibitem [{\citenamefont {{Wildeboer}}\ \emph {et~al.}(2021)\citenamefont
  {{Wildeboer}}, \citenamefont {{Iadecola}},\ and\ \citenamefont
  {{Williamson}}}]{wildeboer2021symmetry}%
  \BibitemOpen
  \bibfield  {author} {\bibinfo {author} {\bibfnamefont {J.}~\bibnamefont
  {{Wildeboer}}}, \bibinfo {author} {\bibfnamefont {T.}~\bibnamefont
  {{Iadecola}}}, \ and\ \bibinfo {author} {\bibfnamefont {D.~J.}\ \bibnamefont
  {{Williamson}}},\ }\href@noop {} {\bibfield  {journal} {\bibinfo  {journal}
  {arXiv e-prints}\ ,\ \bibinfo {eid} {arXiv:2110.05710}} (\bibinfo {year}
  {2021})},\ \Eprint {http://arxiv.org/abs/2110.05710} {arXiv:2110.05710
  [quant-ph]} \BibitemShut {NoStop}%
\bibitem [{\citenamefont {{Tuckett}}\ \emph {et~al.}(2018)\citenamefont
  {{Tuckett}}, \citenamefont {{Bartlett}},\ and\ \citenamefont
  {{Flammia}}}]{tuckett2018ultrahigh}%
  \BibitemOpen
  \bibfield  {author} {\bibinfo {author} {\bibfnamefont {D.~K.}\ \bibnamefont
  {{Tuckett}}}, \bibinfo {author} {\bibfnamefont {S.~D.}\ \bibnamefont
  {{Bartlett}}}, \ and\ \bibinfo {author} {\bibfnamefont {S.~T.}\ \bibnamefont
  {{Flammia}}},\ }\href {\doibase 10.1103/PhysRevLett.120.050505} {\bibfield
  {journal} {\bibinfo  {journal} {\prl}\ }\textbf {\bibinfo {volume} {120}},\
  \bibinfo {eid} {050505} (\bibinfo {year} {2018})},\ \Eprint
  {http://arxiv.org/abs/1708.08474} {arXiv:1708.08474 [quant-ph]} \BibitemShut
  {NoStop}%
\bibitem [{\citenamefont {{Tuckett}}\ \emph {et~al.}(2019)\citenamefont
  {{Tuckett}}, \citenamefont {{Darmawan}}, \citenamefont {{Chubb}},
  \citenamefont {{Bravyi}}, \citenamefont {{Bartlett}},\ and\ \citenamefont
  {{Flammia}}}]{tuckett2019tailoring}%
  \BibitemOpen
  \bibfield  {author} {\bibinfo {author} {\bibfnamefont {D.~K.}\ \bibnamefont
  {{Tuckett}}}, \bibinfo {author} {\bibfnamefont {A.~S.}\ \bibnamefont
  {{Darmawan}}}, \bibinfo {author} {\bibfnamefont {C.~T.}\ \bibnamefont
  {{Chubb}}}, \bibinfo {author} {\bibfnamefont {S.}~\bibnamefont {{Bravyi}}},
  \bibinfo {author} {\bibfnamefont {S.~D.}\ \bibnamefont {{Bartlett}}}, \ and\
  \bibinfo {author} {\bibfnamefont {S.~T.}\ \bibnamefont {{Flammia}}},\ }\href
  {\doibase 10.1103/PhysRevX.9.041031} {\bibfield  {journal} {\bibinfo
  {journal} {Physical Review X}\ }\textbf {\bibinfo {volume} {9}},\ \bibinfo
  {eid} {041031} (\bibinfo {year} {2019})},\ \Eprint
  {http://arxiv.org/abs/1812.08186} {arXiv:1812.08186 [quant-ph]} \BibitemShut
  {NoStop}%
\bibitem [{\citenamefont {{Tuckett}}\ \emph {et~al.}(2020)\citenamefont
  {{Tuckett}}, \citenamefont {{Bartlett}}, \citenamefont {{Flammia}},\ and\
  \citenamefont {{Brown}}}]{tuckett2020faulttolerant}%
  \BibitemOpen
  \bibfield  {author} {\bibinfo {author} {\bibfnamefont {D.~K.}\ \bibnamefont
  {{Tuckett}}}, \bibinfo {author} {\bibfnamefont {S.~D.}\ \bibnamefont
  {{Bartlett}}}, \bibinfo {author} {\bibfnamefont {S.~T.}\ \bibnamefont
  {{Flammia}}}, \ and\ \bibinfo {author} {\bibfnamefont {B.~J.}\ \bibnamefont
  {{Brown}}},\ }\href {\doibase 10.1103/PhysRevLett.124.130501} {\bibfield
  {journal} {\bibinfo  {journal} {\prl}\ }\textbf {\bibinfo {volume} {124}},\
  \bibinfo {eid} {130501} (\bibinfo {year} {2020})},\ \Eprint
  {http://arxiv.org/abs/1907.02554} {arXiv:1907.02554 [quant-ph]} \BibitemShut
  {NoStop}%
\end{thebibliography}
\end{document}